\documentclass[11pt]{article}

\usepackage[final]{acl}

\usepackage{times}
\usepackage{latexsym}

\usepackage[T1]{fontenc}

\usepackage[utf8]{inputenc}

\usepackage{microtype}

\usepackage{inconsolata}

\usepackage{graphicx}

\usepackage{amsmath}
\usepackage{dsfont}
\usepackage{float}
\usepackage{tikz}
\usetikzlibrary{positioning, arrows.meta, shapes, fit, backgrounds, calc}
\usepackage{pgfplots}
\pgfplotsset{compat=1.18}
\usepackage{booktabs}
\usepackage{listings}
\usepackage{xcolor}
\usepackage{array}

\lstdefinestyle{dia}{
  basicstyle=\ttfamily\small,
  frame=single,
  breaklines=true,
  showstringspaces=false,
  columns=flexible,
  keepspaces=true
}
\usepackage[most]{tcolorbox}
\tcbuselibrary{listings}
\lstdefinestyle{trace}{
  basicstyle=\ttfamily\footnotesize,
  breaklines=true,
  showstringspaces=false,
  columns=flexible,
  keepspaces=true
}
\newtcblisting{tracebox}[1]{
  listing only,
  listing options={style=trace},
  enhanced,
  breakable,
  arc=1.5mm,
  boxrule=0.6pt,
  colback=blue!2,
  colframe=blue!30!black,
  colbacktitle=blue!30!black,
  coltitle=white,
  fonttitle=\bfseries\footnotesize,
  title={#1},
  left=2mm, right=2mm, top=1mm, bottom=1mm
}

\title{Data Intelligence Agents: Interpreting, Modeling, and Querying Enterprise Data via Autonomous Coding Agents}

\author{
  \textbf{Anoushka Vyas} \quad
  \textbf{Aarushi Dhanuka} \quad
  \textbf{Sina Khoshfetrat Pakazad} \quad
  \textbf{Henrik Ohlsson} \\[3pt]
  C3 AI \\[2pt]
  \texttt{\{anoushka.vyas, aarushi.dhanuka, sina.pakazad, henrik.ohlsson\}@c3.ai}
}

\begin{document}
\maketitle
\addtocontents{toc}{\protect\setcounter{tocdepth}{-2}}
\begin{abstract}
Production data integration is bottlenecked by repeated, lossy handoffs between data owners, engineers, and analysts who must collaboratively discover, structure, and query enterprise data. We present \textbf{D}ata \textbf{I}ntelligence \textbf{A}gents (DIA), a system of three agents (\emph{Data Interpreter}, \emph{Schema Creator}, and \emph{Query Generator}) that compresses this workflow by treating autonomous coding agents (ACAs) as a first-class abstraction: rather than emitting text, the agents generate, execute, validate, and repair concrete artifacts, draw on a shared memory for experience reuse, and surface each for review by domain experts. DIA is deployed in production for enterprise customers. We study the \emph{Query Generator} in depth and evaluate it in fully autonomous mode across seven SQL benchmarks spanning four task categories and four dialects. It matches or surpasses the best published results on all seven, demonstrating that an architecture grounded in execution, built on ACAs and a shared memory, generalizes across the data intelligence workload with adaptation confined to natural-language instructions.

\end{abstract}


\section{Introduction}
\label{sec:intro}

Enterprise data work rarely fails for lack of data; it fails because raw data must be discovered, understood, structured, and queried before it can support analysis. In practice this passes through repeated, lossy handoffs between the data owners who understand what fields mean, the engineers who structure and validate the data, and the analysts who query it. Each handoff loses context and adds latency: a misread field or an implicit business rule forces schema changes, pipeline rework, and query rewrites. The opportunity is to keep the domain experts who understand the data in control while compressing the engineering cycle.

\begin{figure}[t]
\centering
\includegraphics[width=\columnwidth]{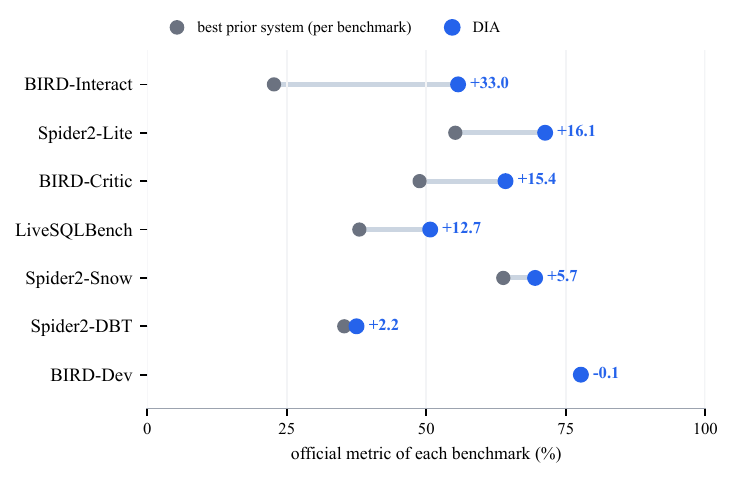}
\caption{DIA against the best prior system on each of the seven SQL benchmarks, ordered by margin. Each benchmark is scored by its official metric.}
\label{fig:results}
\end{figure}

Large language models make each step look tractable in isolation, yet existing systems address fragments of this workflow rather than closing it. Pipeline systems for text-to-SQL chain handcrafted modules, each tuned for one subtask and brittle when the task changes \citep{pourreza2023din, pourreza2024chase}. Specialists trained with reinforcement learning reach high accuracy on a single benchmark but are locked to one dialect and need costly retraining per variant \citep{yang2025mars, wang2025omnisql}. Agentic explorers probe the database live but keep no memory across sessions, restarting from scratch on every query \citep{cao2026apex, lu2025reforce}. SQL agents with persistent memory store and replay past experience, but keep a single store and a narrow evaluation \citep{biswal2025agentsm, yang2026memosql, chu2024expesql, chen2025rubiksql}. Across these approaches the system emits text (queries or critiques) rather than the executable, inspectable artifacts that enterprise data work consumes, and none addresses the upstream understanding and schema construction stages that decide whether the resulting SQL has anything sensible to run against.

\begin{figure*}[t]
\centering
\includegraphics[width=0.95\textwidth]{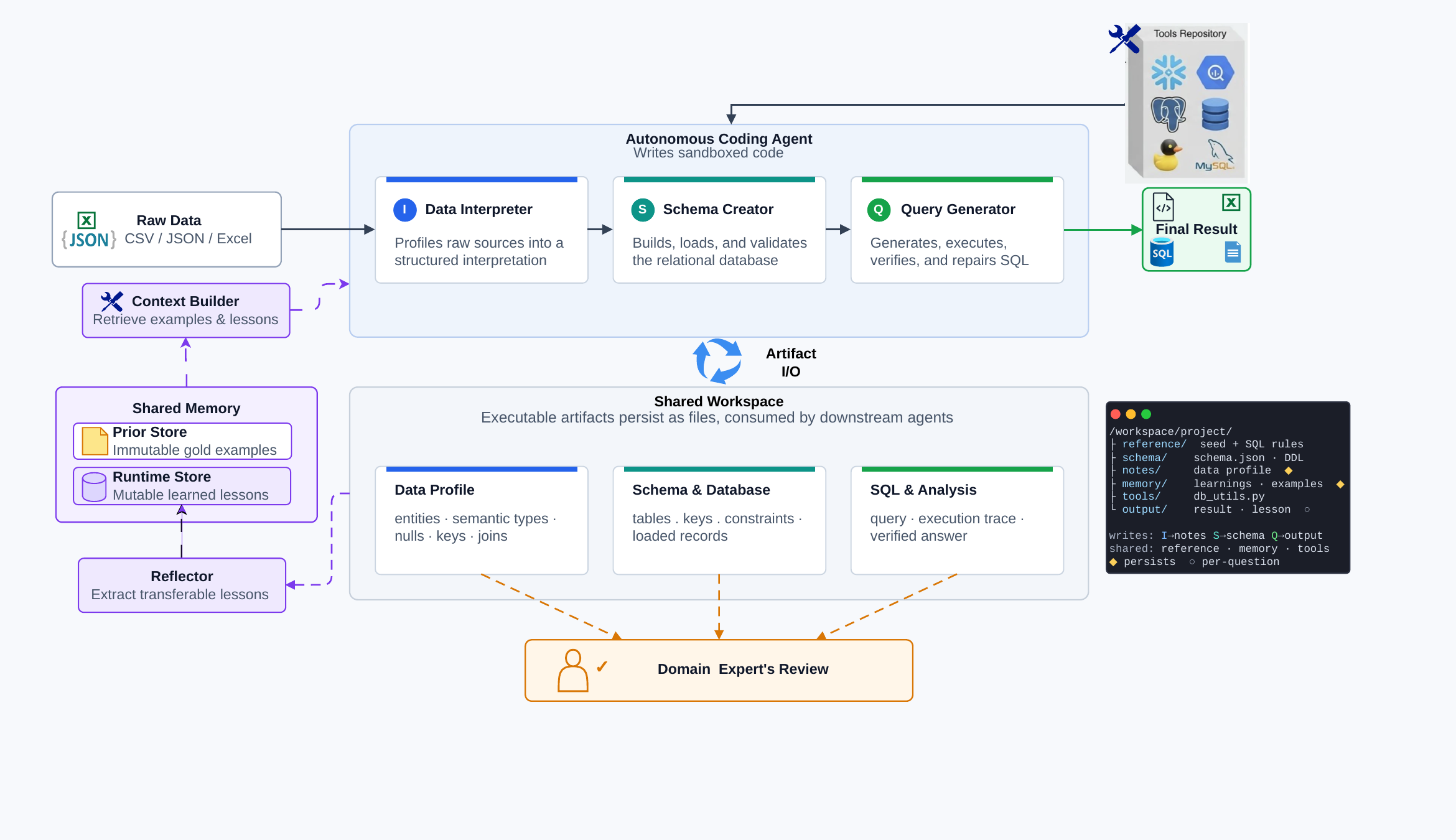}
\caption{The DIA system. A single ACA operating over a shared workspace $W$ realizes three agents (\emph{Data Interpreter}, \emph{Schema Creator}, and \emph{Query Generator}), turning raw data $D$ and a question $q$ into a grounded answer $R$. Each agent reads and writes executable artifacts in $W$; all draw on a shared memory $M$; domain experts review each artifact.}
\label{fig:dia-arch}
\end{figure*}

DIA closes this loop. It directs a single ACA (a coder driven by an LLM in a sandboxed environment) across the discovery, schema construction, and query stages (Figure~\ref{fig:dia-arch}), surfacing each artifact for review by domain experts \citep{wang2024codeact, song2026nemo}. The key contributions are as follows:

\begin{enumerate}
  \item The first system, to our knowledge, to treat the ACA rather than the LLM as the central abstraction for data intelligence: three agents (\emph{Data Interpreter}, \emph{Schema Creator}, and \emph{Query Generator}) realized as a single ACA over a shared workspace that turns raw enterprise data into validated, queryable schemas and grounded answers, replacing lossy text handoffs with inspectable artifacts.
  \item The design of the \emph{Query Generator}: a single generalist agent that handles SQL generation, debugging, conversational interaction, and project completion across four dialects through self-correction grounded in execution and a shared memory for experience reuse, with adaptation confined to natural-language instructions.
  \item A broad empirical study: in fully autonomous mode on seven SQL benchmarks spanning four task categories and four dialects, with a single LLM and no fine-tuning, the \emph{Query Generator} matches or surpasses the best published results on all seven (Figure~\ref{fig:results}).
\end{enumerate}

\section{Related Work}
\label{sec:related}

\paragraph{Text-to-SQL systems.}
Text-to-SQL is a large and active area \citep{hong2025survey}, but the work fragments by task setting. Most systems target single-shot query generation and improve accuracy on it through multi-agent collaboration (MAC-SQL \citep{wang2023macsql}, CHESS \citep{talaei2024chess}), ensemble pipelines (OpenSearch-SQL \citep{xie2025opensearch}, XiYan-SQL \citep{gao2024xiyan}), or component-level advances in schema linking \citep{aclanthology-rosl2025} and decoding \citep{aclanthology-ttdsql2025}. The pipeline, reinforcement-learning, and agentic systems of Section~\ref{sec:intro} address this same setting. Other settings are served by separate, specialized systems: conversational querying (SParC \citep{yu2019sparc}, CoSQL \citep{yu2019cosql}) and declarative querying over heterogeneous data \citep{aclanthology-declarative2025}. The concurrent AgentNLQ \citep{bogdanov2026agentnlq} is the closest, a multi-agent system for general-purpose NL-to-SQL, but it too addresses a single setting. DIA's \emph{Query Generator} instead spans four task categories across four dialects with one agent.

\paragraph{Data understanding and schema generation.}
DIA is a system, not a single SQL model, and its other two agents build on work done so far in isolation. For data understanding, large tabular models profile tables and infer column semantics, types, and relationships (TableGPT2 \citep{su2024tablegpt2}). For schema construction, recent agents build relational schemas from natural language (Text2Schema \citep{wang2025text2schema}) and prepare raw data for analysis (DeepPrep \citep{fan2026deepprep}); new benchmarks measure data agents across the full data intelligence lifecycle, from engineering to analysis (DAComp \citep{lei2026dacomp}). These are standalone tools and evaluations. DIA's \emph{Data Interpreter} and \emph{Schema Creator} instead work as agents in one system, handing validated, executable artifacts to the \emph{Query Generator}.

\paragraph{Generalist and memory agents.}
DIA inherits the generalist coding agent paradigm, where one agent solves many tasks through code execution instead of separate modules for each task (OpenHands-Versa \citep{openhands-versa2025}, CodeAct \citep{wang2024codeact}). It treats ACAs as a first-class abstraction, as does NEMO \citep{song2026nemo} for optimization modeling. DIA also draws on agents that learn from experience: ARIA \citep{aclanthology-aria2025} keeps a knowledge repository that improves over time, and Voyager \citep{wang2023voyager}, Reflexion \citep{shinn2023reflexion}, and ReasoningBank \citep{ouyang2025reasoningbank} accumulate skills, reflections, or reasoning strategies. These build experience for one agent. DIA's three agents instead share a memory and reuse experience across the system.

\section{Methodology}
\label{sec:method}

\subsection{Overview}
\label{sec:method-overview}

A central abstraction in DIA is remote interaction with an ACA, an execution-capable counterpart to a text-only model call. Operating within a sandboxed environment, the ACA generates, executes, inspects, and revises code, so that every output is an executable artifact and admits execution-aware validation \citep{song2026nemo, wang2024codeact}. DIA drives the ACA with natural-language instructions and references to existing workspace artifacts, and receives code, execution traces, and results in return.

DIA is a system of three agents, realized as a single ACA invoked over a shared workspace $W$: a \emph{Data Interpreter} that profiles raw sources into a structured interpretation (Section~\ref{sec:di}); a \emph{Schema Creator} that materializes and validates a relational database from that interpretation (Section~\ref{sec:sc}); and a \emph{Query Generator} that translates natural-language questions into executed SQL (Section~\ref{sec:qg}). The agents communicate through $W$, in which artifacts persist as files, rather than by exchanging text, and each draws on a shared memory $M$ for experience reuse (Section~\ref{sec:memory}); we write $M^{*} \subseteq M$ for the subset retrieved at each invocation. Each artifact is surfaced for review by domain experts.

\subsection{Memory}
\label{sec:memory}

Memory in DIA is artifact-based: because the ACA works in a sandbox, what it carries forward is the concrete artifacts it has produced and validated, schemas, loading and transformation scripts, validation reports, query logs, and prior solutions, rather than textual summaries of them. Within a task, the agents build on the artifacts already in the workspace $W$ (Section~\ref{sec:method-overview}), each consuming what the previous produced. Across tasks, an experience store $M$ retains a reusable subset in three tiers that mirror a memory hierarchy: retrieved examples, an episodic tier of similar past question-and-solution pairs surfaced for the current question; session lessons, conditional rules confirmed on the current database; and cross-session lessons, the long-term semantic subset of those rules that generalize across databases.

Memory is pull-based and verified before use. Items are surfaced only by reference, and the agent reads a body only when it judges it relevant. Nothing changes an answer until a live probe confirms its precondition on the current data, so stale experience is caught by execution rather than propagated. The agent writes memory itself: after answering it reflects and records a conditional rule with its evidence, updating the store
\[
M \leftarrow w(M, a, o),
\]
where $a$ is the artifact just produced and $o$ its observed outcome, and $w$ admits a cross-session lesson only when $o$ bears it out, with no learned or human judge, keeping the mechanism training-free. All three agents share one memory and consume $M^{*}$ through their signatures.

\subsection{Data Interpreter}
\label{sec:di}

Given a collection of heterogeneous raw sources $D = \{d_{1}, \ldots, d_{n}\}$ such as CSV, JSON, and Excel files, the \emph{Data Interpreter} writes and executes profiling code that produces a structured interpretation $P$:
\[
\mathcal{I}: (D, M^*) \to P.
\]
Rather than describe $D$ in natural language, the ACA derives $P$ entirely from executed code, and the resulting artifact is what downstream agents consume. $P$ records, for each source, the inferred schema with column names and semantic types; per-column value distributions, null statistics, and pattern observations; candidate primary and foreign keys; likely join paths across sources; and data-quality observations that require intervention.

\subsection{Schema Creator}
\label{sec:sc}

Given the interpretation $P$ and the raw sources $D$, the \emph{Schema Creator} generates and executes loading and validation code that materializes a working database:
\[
\mathcal{S}: (P, D, M^*) \to (\Sigma, \beta),
\]
where $\Sigma = (T, K, C)$ specifies the tables $T$ with their columns and types, the key constraints $K$ (each table's primary key and the foreign keys linking tables), and the integrity constraints $C$ (column-level rules the data must satisfy, such as not-null and value-range checks), and $\beta$ is $\Sigma$ instantiated as physical tables and populated with the records in $D$. The ACA works under a load-first-normalize-second discipline: staging tables ingest every record from $D$ with provenance fields recording source file and load timestamp; refined tables and views apply typing and structure on top. The ACA then validates $(\Sigma, \beta)$ along four axes: (i) row-count reconciliation between sources and $\beta$; (ii) column coverage, requiring every source column to be carried through and any rename to be recorded; (iii) key validity, checking primary-key uniqueness and foreign-key referential integrity; and (iv) load integrity, routing records that cannot be ingested to per-table reject buffers rather than dropping them silently. A set of test queries $\tau$ is executed against $\beta$; the schema is accepted only when ingestion succeeds and every query in $\tau$ executes as expected. Alongside $\beta$, the ACA emits a schema manifest enumerating $T$, $K$, and column mappings, and a validation report summarizing the four checks.

\subsection{Query Generator}
\label{sec:qg}

Given a natural-language question $q$ and the database $(\Sigma, \beta)$, the \emph{Query Generator} writes and executes SQL to answer it:
\[
\mathcal{Q}: (q, \Sigma, \beta, M^*) \to y,
\]
where $y$ is a \texttt{SELECT} statement for analytical questions or a DDL/DML statement for modification tasks. The ACA accesses $\beta$ in read-only mode for analytical questions. Generation is execution-grounded and proceeds in four phases, with the ACA writing and executing SQL throughout.

\paragraph{Shape declaration.}
Before generating $y$, the ACA derives from $q$ an expected result shape
\[
\kappa = (C_{\kappa},\, g_{\kappa},\, o_{\kappa},\, f_{\kappa}),
\]
where $C_{\kappa}$ is the column list implied by $q$, $g_{\kappa}$ the row granularity (one row per entity, per group, per time bucket, and so on), $o_{\kappa}$ the ordering specification, and $f_{\kappa}$ the filter conjunction extracted from $q$. For modification tasks, $\kappa$ specifies the target objects and the intended post-condition on $\beta$.

\paragraph{Schema exploration.}
The ACA executes lightweight probe queries against $\Sigma$ and $\beta$ to confirm that join keys exist, sample representative column values to fix their format, and verify cardinality assumptions, rather than inferring structure from column names alone.

\paragraph{Generation and execution.}
The ACA produces a candidate query
\[
y = \mathcal{G}(q, \Sigma, M^{*}, \kappa)
\]
conditioned on the question, schema, retrieved memory, and declared shape, and executes it to obtain $R = \mathrm{exec}(y, \beta)$.

\paragraph{Self-verification.}
The agent does not treat the first query it writes as final. It checks the result against the declared shape with its own indicator
\[
V(R, \kappa) = \begin{cases}
1 & \text{if } R \text{ satisfies } \kappa,\\
0 & \text{otherwise},
\end{cases}
\]
evaluated componentwise against $C_{\kappa}$, $g_{\kappa}$, $o_{\kappa}$, and $f_{\kappa}$. For modification tasks, $V$ checks that the intended post-condition holds in $\beta$. This check is the agent's own and is computed from execution rather than supplied by an external verifier or human: when $V(R, \kappa) = 0$, the agent diagnoses the gap, revises $y$, and re-executes within the same pass before emitting an answer. The procedure is independent of task category and SQL dialect; only the grammar of $\kappa$ varies across them.

\section{Evaluation}
\label{sec:eval}

\subsection{Setup}
\label{sec:eval-setup}

We evaluate the \emph{Query Generator} on seven public SQL benchmarks: BIRD-Dev \citep{li2023bird}, BIRD-Critic \citep{birdcritic}, LiveSQLBench \citep{livesqlbench}, BIRD-Interact \citep{birdinteract}, and the Spider2 family \citep{lei2024spider2} (Spider2-Lite, Spider2-Snow, and Spider2-DBT). Together they comprise 4{,}187 instances spanning four task categories and four SQL dialects: generation (BIRD-Dev, LiveSQLBench, Spider2-Lite, Spider2-Snow), debugging (BIRD-Critic), conversational interaction (BIRD-Interact), and dbt project completion (Spider2-DBT), across SQLite, PostgreSQL, Snowflake, and DuckDB. Several benchmarks contain finer task categories, such as data modification in LiveSQLBench and personalization in BIRD-Critic, which we break down in Appendix~\ref{app:perdb}; full dataset details are given in Appendix~\ref{app:exp}.

All experiments use a unified system configuration: OpenHands, powered by Claude Sonnet 4.5 with no fine-tuning, acts as the ACA, while o3 serves as the user simulator in BIRD-Interact's conversational protocol. Customization for each benchmark is confined to a standing seed file and the per-question prompt scaffolding (Appendix~\ref{sec:appA}), and every run is fully autonomous with no human intervention. Implementation details are given in Appendix~\ref{app:impl}. The shared memory examined in Section~\ref{sec:eval-memory} retrieves its episodic examples from the BIRD train split, indexed offline and disjoint from the BIRD-Dev evaluation set, and is detailed in Appendix~\ref{app:memory}.

For each benchmark we compare against the best available prior results (Appendix~\ref{app:leaderboards}). Systems with an accompanying publication are cited in Table~\ref{tab:main}, and the remainder are listed by name. The primary metric is each benchmark's official one: execution accuracy throughout, except task success rate on BIRD-Interact and database-match accuracy on Spider2-DBT. Additional official metrics and their definitions are given in Appendices~\ref{app:perdb} and~\ref{app:metrics}. Reported execution accuracy on BIRD-Dev varies by a small margin across works, owing to differences in evaluation harnesses and to noise and periodic corrections in the benchmark's gold queries \citep{wretblad2024noise}. We release granular per-instance results via HuggingFace\footnote{\url{https://huggingface.co/datasets/c3aiia3c/dia-emnlp2026}}.

\subsection{Main results}
\label{sec:eval-main}

Table~\ref{tab:main} summarizes performance across the seven benchmarks, comparing the \emph{Query Generator} against the strongest available agent-based and training-based systems. Overall, it achieves strong and consistent performance, matching or surpassing the best published result on all seven benchmarks, by large margins on several.

\begin{table*}[t]
\centering
\footnotesize
\setlength{\tabcolsep}{4pt}
\begin{tabular}{l l l l c}
\toprule
\textbf{Benchmark} & \textbf{Dialect} & \textbf{Task} & \textbf{System} & \textbf{Score (\%)} \\
\midrule
BIRD-Dev      & SQLite     & Generation               & Agentar-Scale-SQL~\citep{agentarscalesql} & 74.9 \\
              &            &                          & CHASE-SQL~\citep{pourreza2024chase}       & 74.9 \\
              &            &                          & \textbf{DIA}                       & 77.7 \\
              &            &                          & MARS-SQL~\citep{yang2025mars}             & \textbf{77.8} \\
\midrule
BIRD-Critic   & SQLite     & Debugging                & Claude Opus 4.6                           & 46.2 \\
              &            &                          & BIRD-Talon-14B                            & 48.0 \\
              &            &                          & Gemini 3.1 Pro Preview                    & 48.8 \\
              &            &                          & \textbf{DIA}                       & \textbf{64.2} \\
\midrule
LiveSQLBench  & PostgreSQL & Generation               & OpenHands $+$ Kimi 2.5                    & 32.2 \\
              &            &                          & OpenHands $+$ Claude Sonnet 4.5           & 35.2 \\
              &            &                          & OpenHands $+$ Claude Opus 4.6             & 38.0 \\
              &            &                          & \textbf{DIA}                       & \textbf{50.7} \\
\midrule
BIRD-Interact & PostgreSQL & Conversational           & Claude Opus 4.6                           & 17.5 \\
              &            &                          & MERIT~\citep{merit} $+$ Claude Opus 4.6   & 20.8 \\
              &            &                          & MERIT~\citep{merit} $+$ GPT-5.4           & 22.7 \\
              &            &                          & \textbf{DIA}                       & \textbf{55.7} \\
\midrule
Spider2-Lite  & Snowflake, SQLite & Generation        & DSR-SQL~\citep{dsrsql}                     & 46.8 \\
              &            &                          & AutoLink~\citep{autolink} $+$ DeepSeek-R1          & 52.3 \\
              &            &                          & ReFoRCE~\citep{lu2025reforce} $+$ o3      & 55.2 \\
              &            &                          & \textbf{DIA}                       & \textbf{71.3} \\
\midrule
Spider2-Snow  & Snowflake  & Generation               & APEX-SQL~\citep{cao2026apex}              & 53.0 \\
              &            &                          & ReFoRCE~\citep{lu2025reforce} $+$ o3      & 62.9 \\
              &            &                          & DSR-SQL~\citep{dsrsql} $+$ DeepSeek-R1             & 63.8 \\
              &            &                          & \textbf{DIA}                       & \textbf{69.5} \\
\midrule
Spider2-DBT   & DuckDB     & Project completion       & Spider-Agent~\citep{lei2024spider2} $+$ o1-preview & 13.2 \\
              &            &                          & Spider-Agent $+$ Claude 3.7 Sonnet        & 14.7 \\
              &            &                          & Spider-Agent-DBT $+$ GPT-5.4              & 35.3 \\
              &            &                          & \textbf{DIA}                       & \textbf{37.5} \\
\bottomrule
\end{tabular}
\caption{DIA against the strongest published baselines across seven SQL benchmarks. Score is each benchmark's official primary metric (higher is better): execution accuracy in all cases except task success rate (BIRD-Interact) and database-match accuracy (Spider2-DBT). Each benchmark's primary and additional metrics are described in detail in Appendix~\ref{app:metrics}.}
\label{tab:main}
\end{table*}

The gains are largest on the tasks where prior systems struggle most: +33.0 points on BIRD-Interact (conversational interaction), +16.1 on Spider2-Lite, +15.4 on BIRD-Critic (debugging), and +12.7 on LiveSQLBench. Appendix~\ref{sec:appD} traces where the BIRD-Interact margin comes from, with worked interaction traces and an interaction-time scaling analysis. On the most competitive benchmarks the lead narrows but holds: +5.7 on Spider2-Snow and +2.2 on Spider2-DBT, where the agent edits a real dbt repository rather than emitting a single query and the strongest prior system is built on GPT-5.4. On BIRD-Dev, the most saturated benchmark, where the field is clustered within a point, DIA is level with the strongest published result, MARS-SQL~\citep{yang2025mars}, an RL-trained specialist (77.7 vs.\ 77.8). One model and scaffold thus serve four task categories and four dialects, with adaptation confined to natural-language standing instructions.

The per-category breakdown (Appendix~\ref{app:perdb}) shows two consistent patterns beneath the headline scores. DIA is strongest on the more structured task variants: BIRD-Critic Management (78.7) and LiveSQLBench Modification (66.3) both exceed the same benchmark's pure-query slice, because modification tasks reward the agent's habit of declaring the target object and validating it by execution before answering. It is weakest on high-level questions that name a composite metric without spelling out its formula (41.6 on LiveSQLBench, 47.6 on BIRD-Interact), where the agent must either decompose the metric or ask. On BIRD-Interact, the phase-2 conditional pass rate of 86.8 shows that once a correct phase-1 query lands, follow-ups are almost always answered correctly.

\subsection{Component ablation}
\label{sec:eval-ablation}

To separate the contribution of the system from that of the underlying model, we ablate one component at a time on the full BIRD-Dev set (1{,}534 questions), holding the model and the agent loop fixed (Table~\ref{tab:ablation}). Removing shape declaration and removing execution-based self-verification each cost 7.5 points; these are two halves of one output-contract discipline (Appendix~\ref{sec:appA}), so disabling either one removes it. Removing the shared memory costs 2.6 points. Reducing the system to a bare OpenHands agent on the same model, given only the schema and the question, costs 14.1 points, a second system-versus-model control that mirrors the LiveSQLBench gap in Table~\ref{tab:main} (35.2 to 50.7). The clarification policy is evaluated on BIRD-Interact, whose interactive protocol depends on it: removing it drops task success from 55.7 to 14.5, a 41.2-point fall (Appendix~\ref{sec:appD}). Across benchmarks, then, it is these components rather than the backbone that drive the gains.

\begin{table*}[t]
\centering
\small
\begin{minipage}[t]{0.48\textwidth}
\centering
\setlength{\tabcolsep}{5pt}
\begin{tabular}{@{}l c c@{}}
\toprule
\textbf{Configuration} & \textbf{Score (\%)} & \textbf{$\Delta$} \\
\midrule
Full DIA                              & 77.7 & --- \\
\quad No shared memory                & 75.1 & --2.6 \\
\quad No self-verification            & 70.2 & --7.5 \\
\quad No shape declaration            & 70.2 & --7.5 \\
\quad Bare OpenHands agent            & 63.6 & --14.1 \\
\midrule
Full DIA, mean of 3 runs              & 76.9 $\pm$ 0.8 & --- \\
\bottomrule
\end{tabular}
\caption{Component ablation on the full BIRD-Dev set (1{,}534 questions), removing one component at a time on the same model and instances, plus the mean and standard deviation across three independent full runs. The clarification policy is ablated separately on BIRD-Interact, dropping from 55.7 to 14.5, a fall of 41.2 points, whose protocol requires it.}
\label{tab:ablation}
\end{minipage}
\hfill
\begin{minipage}[t]{0.48\textwidth}
\centering
\setlength{\tabcolsep}{5pt}
\begin{tabular}{@{}l l@{}}
\toprule
\textbf{Metric} & \textbf{Mean} \\
\midrule
Cost / question          & USD 0.10 (range 0.03--0.20) \\
Input tokens / question  & 145k (94\% cached) \\
Output tokens / question & 1.6k \\
Runtime, median          & 38\,s \\
Runtime, 90th percentile & 68\,s \\
\bottomrule
\end{tabular}
\caption{Per-question cost, token usage, and latency for DIA on BIRD-Dev, averaged over the full run; the cost range is across databases. Runtime is the distribution over questions; the 90th percentile bounds all but the slowest one in ten.}
\label{tab:cost}
\end{minipage}
\end{table*}

This ablation also maps design to benchmark gains. Shape declaration and self-verification, the output-contract discipline ablated above, underpin the debugging and modification gains on BIRD-Critic (+15.4) and LiveSQLBench (+12.7) in Table~\ref{tab:main}, since both tasks reward validating a candidate object or edit against the database before returning it. The clarification policy accounts for most of the +33.0-point margin on BIRD-Interact, whose conversational protocol rewards resolving ambiguity rather than guessing. Shared memory contributes more evenly across benchmarks, since the rules it accumulates concern joins, aggregation, and output convention, failure modes that recur regardless of task type (Section~\ref{sec:eval-memory}).

\subsection{Cost, latency, and variance}
\label{sec:eval-cost}

We report per-question cost and latency for DIA on BIRD-Dev, measured over the full run (Table~\ref{tab:cost}). A question costs about USD 0.10 on average, ranging from 0.03 to 0.20 across databases, and processes about 145k input tokens (of which roughly 94\% are served from cache) and about 1.6k output tokens. End-to-end runtime has a median of 38 seconds; the 90th percentile, meaning all but the slowest one in ten questions, is 68 seconds. Since questions are independent, throughput scales with parallel workers.

To quantify run-to-run variability under the non-determinism of the underlying agent, we run BIRD-Dev three times, obtaining 77.7, 76.1, and 77.0, a mean of 76.9 with a standard deviation of 0.8 (Table~\ref{tab:ablation}). The 0.1-point gap to the strongest published result in Table~\ref{tab:main} is thus well within run-to-run noise, and within the documented annotation noise of the benchmark \citep{wretblad2024noise}.

\subsection{Learned Rules}
\label{sec:eval-memory}

On BIRD-Dev, for example, DIA accumulates a small store of conditional rules of a few recurring kinds, each interpretable and grounded in a check on the data. This accumulation introduces no test leakage: the agent never sees gold answers or a grading signal, every rule is distilled from its own execution observations and re-verified on the live database before it can change an answer, and the episodic examples it retrieves come from the disjoint BIRD train split. Rules form in two stages: a rule begins as a concrete within-database observation and is promoted to a cross-database rule only when later questions bear it out. On \texttt{california\_schools}, for instance, the agent counting schools through a one-to-many join saw \texttt{COUNT(*)} return 9{,}977 for one school (its fact-row count) but \texttt{COUNT(DISTINCT CDSCode)} return 1. The rule promoted from this is to count entities through a repeating join with \texttt{COUNT(DISTINCT pk)}, not \texttt{COUNT(*)}. The rule kinds it records span joins, aggregation, and output convention, the same recurring failure structures the error analysis (Section~\ref{sec:eval-errors}) finds behind most wrong answers. Appendix~\ref{app:memory} gives representative rules with their evidence, both stages of this promotion, the full three-tier store, and a worked trace of a session lesson redirecting a later answer.

\subsection{Error analysis}
\label{sec:eval-errors}

Almost every failed instance runs to completion and returns a wrong answer rather than failing to execute, so the remaining errors are overwhelmingly semantic rather than syntactic. Appendix~\ref{app:errors} sorts them into three recurring classes: reasoning, output convention, and grounding, with reasoning by far the most frequent. A reasoning failure executes cleanly but answers a subtly different question, most often a wrong join, filter, or formula on an under-specified request.


\section{Conclusion}
\label{sec:conclusion}

We presented DIA, which treats the ACA as the central abstraction for enterprise data intelligence: a single ACA over a shared workspace, invoked as three agents, produces executable artifacts that domain experts review rather than text they must trust. The premise is that one agent grounded in execution can replace a family of specialized systems, and our evaluation supports it: with a single LLM and no fine-tuning, the \emph{Query Generator} matches or surpasses the best published results on all seven benchmarks (Table~\ref{tab:main}), across task categories and dialects that prior work covers with separate, task-specific systems.

Two lessons recur across benchmarking and deployment alike. Execution grounding helps most where an answer can be checked against the database, and least on upstream domain ambiguity that execution cannot resolve; and for a system meant to be reviewed rather than trusted outright, the design choice that matters most is emitting inspectable artifacts rather than natural-language explanations of them. DIA is deployed in production for enterprise customers (Appendix~\ref{app:deploy}), where these same properties, execution-grounded artifacts, human review, and permission-scoped access, provide the auditability such settings require.


\section{Limitations and Future Work}
\label{sec:limitations}

DIA trades computation for reliability. Each question runs through an iterative loop of generation, execution, and verification, with conversational tasks adding multi-turn interaction, so mean time per question ranges from under a minute on single-shot generation tasks to roughly ten minutes on multi-turn conversational tasks; Section~\ref{sec:eval-cost} reports the full BIRD-Dev cost and latency distribution. This is acceptable when answers feed durable artifacts but may be prohibitive for interactive or high-throughput use. Caching exploration, parallelizing independent runs, and distilling routine patterns into cheaper components would reduce it.

Verification in DIA is execution-grounded but not semantic. The agent checks an executed result against the result shape it has itself derived from the question, so when it misreads intent, the query and the check inherit the same misreading and a wrong answer passes. The residual failures in Section~\ref{sec:eval-errors} concentrate there: join, filter, and formula reasoning on under-specified questions. Engaging the semantics of the question, rather than the shape of its answer, is the clearest path to closing the gap.

Finally, the evaluation is broad in tasks and dialects but deliberately narrow elsewhere: one of the three agents, a single backbone, a simulated rather than human user on conversational tasks, and memory examined only qualitatively. All results use one LLM (Claude Sonnet 4.5) with no fine-tuning, so we do not yet establish that the gains transfer across backbones; the ablation in Section~\ref{sec:eval-ablation} bounds the single-model risk, since DIA's margin over a bare agent on the same model and substrate (14.1 points) indicates the system, not the model, carries the result, but a study across backbones remains future work. We plan to widen each dimension: benchmarking the \emph{Data Interpreter} and \emph{Schema Creator} on data preparation and schema generation, testing sensitivity across LLMs such as GPT and Gemini, and studying real users. Memory is a further direction: experience accumulates as artifacts and rules in the workspace, and organizing it with graph-structured links rather than as files would let the system mine that experience more effectively.

\bibliography{custom}

\clearpage
\appendix
\onecolumn

\addtocontents{toc}{\protect\setcounter{tocdepth}{2}}
\renewcommand{\contentsname}{Appendix Contents}
\tableofcontents
\clearpage

\section{Datasets}
\label{app:exp}
\label{app:metrics}

\subsection{Descriptions}
\label{app:exp-benchmarks}

Table~\ref{tab:appA-datasets} summarizes the seven benchmarks. All are public, and we evaluate their official splits.

\begin{table*}[ht]
\centering
\small
\setlength{\tabcolsep}{4pt}
\begin{tabular}{l r r l l l l}
\toprule
\textbf{Benchmark} & \textbf{N} & \textbf{DBs} & \textbf{Dialect} & \textbf{Task} & \textbf{Composition} & \textbf{Metric} \\
\midrule
BIRD-Dev      & 1{,}534 &  11 & SQLite            & generation         & by difficulty: simple 922,         & EX \\
              &         &     &                   &                    & moderate 460, challenging 152      &    \\
\addlinespace
BIRD-Critic   &   500   &  15 & SQLite            & debugging          & by category: query 284,            & EX \\
              &         &     &                   &                    & personalization 141, management 75 &    \\
\addlinespace
LiveSQLBench  &   600   &  22 & PostgreSQL        & generation         & by category: query 410,            & EX \\
              &         &     &                   &                    & modification 190                   &    \\
              &         &     &                   &                    & by difficulty: high-level 286,     &    \\
              &         &     &                   &                    & non-high-level 314                 &    \\
\addlinespace
BIRD-Interact &   600   &  22 & PostgreSQL        & conversational     & by category: query 410,            & SR \\
              &         &     &                   &                    & management 190                     &    \\
              &         &     &                   &                    & by difficulty: high-level 286,     &    \\
              &         &     &                   &                    & low-level 314                      &    \\
\addlinespace
Spider2-Lite  &   342   &  88 & SQLite, Snowflake & generation         & by dialect: SQLite 135,            & EX \\
              &         &     &                   &                    & Snowflake 207                       &    \\
\addlinespace
Spider2-Snow  &   547   & 152 & Snowflake         & generation         & ---                                & EX \\
\addlinespace
Spider2-DBT   &    64   &  64 & DuckDB            & project completion & ---                                & DM \\
\bottomrule
\end{tabular}
\caption{The seven benchmarks: instance counts, databases, dialects, tasks, composition, and primary metrics (EX: execution accuracy; SR: task success rate; DM: database match; defined in Appendix~\ref{app:exp-metrics}). The composition categories are defined in the per-benchmark descriptions of Appendix~\ref{app:exp-benchmarks}.}
\label{tab:appA-datasets}
\end{table*}

\paragraph{BIRD-Dev \citep{li2023bird}.}
The development split of BIRD: 1{,}534 natural language questions over 11 SQLite databases spanning 37 professional domains, each with optional external knowledge evidence. Difficulty tiers are assigned by the benchmark authors and reflect the complexity of the required SQL. We use the development split because the test split is hidden; as noted in Section~\ref{sec:eval-setup}, reported figures on this split vary by a small margin across works \citep{wretblad2024noise}.

\paragraph{BIRD-Critic \citep{birdcritic}.}
The BIRD-Critic-SQLite release: 500 user issues, each consisting of a problem statement and (usually) a buggy SQL fragment that the agent must diagnose and fix. Query issues debug a failing analytical query; personalization issues tailor a query to a user-specific requirement beyond the literal bug; management issues repair statements that modify the schema or data. Instances are graded by benchmark-shipped test cases against the corrected query or statement.

\paragraph{LiveSQLBench \citep{livesqlbench}.}
The LiveSQLBench-Base-Full v1 release: 600 instances over 22 full-scale PostgreSQL databases with hierarchical external knowledge documents. Query instances require analytical SELECTs; modification instances require DDL or DML graded by test cases. High-level instances pose the question through knowledge-base concepts, often composite metrics whose definitions the agent must resolve, while non-high-level instances state the requested computation directly.

\paragraph{BIRD-Interact \citep{birdinteract}.}
The full split of BIRD-Interact: 600 two-phase conversational instances over the same 22 PostgreSQL databases. The user is played by an LLM simulator that resolves ambiguities the benchmark deliberately injects; the agent acts through an ASK/SUBMIT protocol with a per-phase turn budget. An instance succeeds only if both phases pass. Query instances request analytical SELECTs and management instances request schema or data changes; the high-level and low-level split mirrors LiveSQLBench's, separating questions posed through knowledge-base concepts from directly stated ones. Phase-2 follow-ups build on the accepted Phase-1 answer and span five types: aggregation (120) summarizes the Phase-1 result; attribute change (114) adds, removes, or replaces reported columns; constraint change (49) alters the conditions of Phase 1; result-based follow-ups (266) pose a new question that depends on the values Phase 1 returned; and topic pivot (51) shifts to a related question on the same database.

\paragraph{Spider2 \citep{lei2024spider2}.}
The Spider2 family targets enterprise-scale warehouses. Spider2-Lite and Spider2-Snow are single-query generation tasks against large schemas; Spider2-DBT asks the agent to complete a dbt project so that the final built database matches gold.

\subsection{Metrics}
\label{app:exp-metrics}

For instance $i$ of a benchmark with $N$ instances, following the notation of Section~\ref{sec:method}, let $y_i$ be the SQL the agent submits and $R_i = \mathrm{exec}(y_i, \beta_i)$ its executed result on the instance database $\beta_i$; let $y^{*}_i$ and $R^{*}_i$ denote the benchmark's gold query and its result.

\paragraph{Execution accuracy \citep{li2023bird}.}
The fraction of instances whose executed result matches gold,
\[
\mathrm{EX} = \frac{1}{N} \sum_{i=1}^{N} \mathds{1}\!\left[ R_i = R^{*}_i \right],
\]
where equality is taken under each benchmark's official comparison rules: row order only where the question requires it, per-instance decimal precision, and set semantics otherwise; where a benchmark provides several acceptable gold results, matching any one suffices. BIRD-Critic and LiveSQLBench replace the direct equality with benchmark-shipped test cases, so the indicator is one exactly when every test case passes against $R_i$ or against the post-modification database state.

\paragraph{Soft-F1 \citep{li2023bird}.}
Rows of $R_i$ and $R^{*}_i$ are aligned in order, and each aligned pair contributes the fraction of its cells that match to $\mathrm{TP}_i$, with the unmatched fractions counted toward $\mathrm{FP}_i$ (prediction-only) and $\mathrm{FN}_i$ (gold-only); rows without a counterpart count wholly toward $\mathrm{FP}_i$ or $\mathrm{FN}_i$. With per-instance precision $P_i = \mathrm{TP}_i / (\mathrm{TP}_i + \mathrm{FP}_i)$ and recall $C_i = \mathrm{TP}_i / (\mathrm{TP}_i + \mathrm{FN}_i)$:
\[
\text{soft-F1} = \frac{1}{N} \sum_{i=1}^{N} \frac{2\, P_i\, C_i}{P_i + C_i}.
\]
The metric grants partial credit for partially correct rows that execution accuracy scores as outright failures.

\paragraph{Valid Efficiency Score \citep{li2023bird}.}
For each correctly answered instance, the gold and predicted execution times give a ratio $\tau_i = E(y^{*}_i) / E(y_i)$, which the benchmark's evaluation code maps to a step reward $\rho(\tau)$: $1.25$ for $\tau \geq 2$, $1$ for $\tau \in [1, 2)$, $0.75$ for $\tau \in [0.5, 1)$, $0.5$ for $\tau \in [0.25, 0.5)$, and $0.25$ below:
\[
\mathrm{VES} = \frac{1}{N} \sum_{i=1}^{N} \mathds{1}\!\left[ R_i = R^{*}_i \right] \cdot \sqrt{\rho(\tau_i)},
\]
so a correct query at least as fast as gold earns full or bonus credit, and an incorrect one earns none.

\paragraph{Task success rate \citep{birdinteract}.}
Let $s^{(1)}_i, s^{(2)}_i \in \{0, 1\}$ indicate that phase 1 and phase 2 of instance $i$ pass within their turn budgets. An instance succeeds only when both do:
\[
\mathrm{SR} = \frac{1}{N} \sum_{i=1}^{N} s^{(1)}_i\, s^{(2)}_i.
\]

\paragraph{Normalized reward \citep{birdinteract}.}
The benchmark's phase-weighted score, which grants partial credit for a correct first phase using the official phase weights:
\[
\text{reward} = \frac{1}{N} \sum_{i=1}^{N} \left( 0.7\, s^{(1)}_i + 0.3\, s^{(2)}_i \right),
\]
reported in percentage points.

\paragraph{Database match \citep{lei2024spider2}.}
For Spider2-DBT, building the agent's completed dbt project produces a database $\beta'_i$, whose benchmark-specified tables and columns are compared with the gold build $\beta'^{*}_i$:
\[
\mathrm{DM} = \frac{1}{N} \sum_{i=1}^{N} \mathds{1}\!\left[ \beta'_i = \beta'^{*}_i \right].
\]

\section{Additional Results}
\label{sec:appE}
\label{app:perdb}

The main paper reports a single headline score per benchmark
(Table~\ref{tab:main}). This appendix reports the finer slices:
category and phase breakdowns, difficulty and dialect breakdowns,
and per-database results.

\subsection{Category and phase breakdowns}

Table~\ref{tab:appE-breakdown} consolidates the sub-scores of the
three benchmarks that ship finer task categories. Two patterns
recur. Structured task variants score above their pure-query
counterparts (BIRD-Critic management 78.7 against query 64.4,
LiveSQLBench modification 66.3 against query 43.4), because
modification tasks name their target objects and admit direct
validation by execution. High-level questions cost fifteen
to seventeen points on both PostgreSQL benchmarks (41.6 against 58.9
on LiveSQLBench, 47.6 against 63.1 on BIRD-Interact): resolving a
knowledge-base concept into the right formula is harder than
implementing a stated computation. On BIRD-Interact, phase~1 is
the bottleneck (64.2 pass rate), while the conditional phase-2
rate of 86.8 shows that follow-ups rarely fail once a correct
base answer exists.

Table~\ref{tab:appE-followup} decomposes BIRD-Interact's second
phase by follow-up type. Mechanical transformations of the
accepted answer are the easiest: aggregation follow-ups pass
95.9\% of the time conditioned on phase~1. Result-based
follow-ups, which depend on the values the first answer returned,
have the lowest conditional rate (82.4), and topic pivots score
highest unconditionally (68.6) because they behave like fresh
questions on a database the agent has already explored.

\begin{table}[H]
\centering
\small
\setlength{\tabcolsep}{5pt}
\begin{tabular}{l r r r}
\toprule
\textbf{Slice} & \textbf{Total} & \textbf{Correct} & \textbf{Acc.} \\
\midrule
\multicolumn{4}{l}{\emph{BIRD-Critic (by category)}} \\
\quad Query           & 284 & 183 & 64.4 \\
\quad Personalization & 141 &  79 & 56.0 \\
\quad Management      &  75 &  59 & 78.7 \\
\midrule
\multicolumn{4}{l}{\emph{LiveSQLBench (by category and difficulty)}} \\
\quad Query           & 410 & 178 & 43.4 \\
\quad Modification    & 190 & 126 & 66.3 \\
\quad High-level      & 286 & 119 & 41.6 \\
\quad Non-high-level  & 314 & 185 & 58.9 \\
\midrule
\multicolumn{4}{l}{\emph{BIRD-Interact (by phase, category, difficulty)}} \\
\quad Phase-1 pass             & 600 & 385 & 64.2 \\
\quad Phase-2 (conditional)    & 385 & 334 & 86.8 \\
\quad Query                    & 410 & 240 & 58.5 \\
\quad Management               & 190 &  94 & 49.5 \\
\quad High-level               & 286 & 136 & 47.6 \\
\quad Low-level                & 314 & 198 & 63.1 \\
\bottomrule
\end{tabular}
\caption{Consolidated per-category, per-phase, and per-difficulty
breakdown for the three benchmarks that report sub-scores.
BIRD-Interact also reports a normalized reward of 61.6.}
\label{tab:appE-breakdown}
\end{table}

\begin{table}[H]
  \centering
  \small
  \begin{tabular}{l r r r r}
    \toprule
    Follow-up type & N & Passed & Success (\%) & Cond.\ (\%) \\
    \midrule
    aggregation & 120 & 70 & 58.3 & 95.9 \\
    attribute change & 114 & 72 & 63.2 & 86.7 \\
    constraint change & 49 & 21 & 42.9 & 87.5 \\
    result-based & 266 & 136 & 51.1 & 82.4 \\
    topic pivot & 51 & 35 & 68.6 & 87.5 \\
    \bottomrule
  \end{tabular}
  \caption{BIRD-Interact results by phase-2 follow-up type
    (the types are defined in Appendix~\ref{app:exp-benchmarks}).
    Cond.\ is the pass rate among instances whose Phase~1 passed;
    it is high in part because the runner echoes the accepted
    Phase-1 SQL into the Phase-2 prompt, anchoring the follow-up
    on the accepted base rather than a rewrite (see
    Appendix~\ref{sec:appD}).}
  \label{tab:appE-followup}
\end{table}

\subsection{Difficulty and dialect breakdowns}

Table~\ref{tab:appE-birddev} reports BIRD-Dev by difficulty
together with its additional official metrics. All three metrics
fall monotonically with difficulty, and soft-F1 sits above
execution accuracy at every tier with a gap that widens from 1.6
points on simple questions to 3.1 on challenging ones: harder
questions are increasingly answered almost correctly, with
partial row overlap that the strict metric scores as failure.

Table~\ref{tab:appE-spider2lite} splits Spider2-Lite by dialect.
The SQLite subset scores seven points above the Snowflake subset
(75.6 against 68.6), reflecting the larger schemas and
semi-structured columns of the warehouse side; the gap is modest,
consistent with the dialect robustness the main results show
across benchmarks.

\begin{table}[H]
  \centering
  \small
  \begin{tabular}{l c c c}
    \toprule
    \textbf{BIRD-Dev} & \textbf{EX} & \textbf{soft-F1} & \textbf{VES} \\
    \midrule
    Simple       & 83.2 & 84.8 & 77.4 \\
    Moderate     & 72.2 & 74.2 & 65.8 \\
    Challenging  & 61.2 & 64.3 & 56.6 \\
    \midrule
    Overall      & 77.7 & 79.2 & 71.9 \\
    \bottomrule
  \end{tabular}
  \caption{BIRD-Dev (1{,}534 instances) by difficulty,
    with the benchmark's additional official metrics (soft-F1,
    VES). As noted in the main paper, reported BIRD-Dev
    execution accuracy varies by a small margin across published
    works.}
  \label{tab:appE-birddev}
\end{table}

\begin{table}[H]
  \centering
  \small
  \begin{tabular}{l r r r}
    \toprule
    Dialect & N & Correct & EX (\%) \\
    \midrule
    SQLite & 135 & 102 & 75.6 \\
    Snowflake & 207 & 142 & 68.6 \\
    \midrule
    \textbf{Total} & \textbf{342} & \textbf{244} & \textbf{71.3} \\
    \bottomrule
  \end{tabular}
  \caption{Spider2-Lite results by dialect.}
  \label{tab:appE-spider2lite}
\end{table}

\subsection{Per-database results}

Tables \ref{tab:appE-livesqlbench}, \ref{tab:appE-birdcritic},
\ref{tab:appE-birdinteract}, and \ref{tab:appE-birddev-perdb} report
per-database results for LiveSQLBench, BIRD-Critic, BIRD-Interact,
and BIRD-Dev, each sorted by accuracy. The spread is widest on LiveSQLBench, where
per-database accuracy ranges from 76.5 to 19.2: the top of the
table is well-specified operational data with clean joins, and
the bottom is dominated by domain ambiguity in the knowledge
base. LiveSQLBench and BIRD-Interact share the same 22 databases,
and their rankings largely agree (\texttt{reverse\_logistics}
and \texttt{cybermarket\_pattern} anchor the top of both tables
while \texttt{mental\_health} and \texttt{organ\_transplant}
anchor the bottom), indicating that difficulty is chiefly a
property of the database and its knowledge base rather than of
the interaction protocol. BIRD-Critic is more uniform: among its
databases with at least ten issues, accuracy stays within a band
from 54.8 to 75.0, suggesting debugging difficulty depends less
on the domain than question answering does.

For the Spider2 family we report the distribution of per-database
outcomes instead (Table~\ref{tab:appE-spider2dist}): its
databases hold only a few questions each, so individual
per-database accuracies are coarse, and full tables over its 88,
152, and 64 databases would add pages without adding signal.
Across Spider2-Lite and Spider2-Snow, DIA fully solves 99 of the
240 databases and is shut out on 40, so the headline accuracy
reflects broad competence rather than a few concentrated wins.
Spider2-DBT is the hardest member, with 40 of 64 projects
incomplete, consistent with project completion being the hardest
task in Table~\ref{tab:main}. Per-database and per-project
outcomes are part of the released per-instance results.

\begin{table}[H]
  \centering
  \small
  \begin{tabular}{l r r r r}
    \toprule
    Database & Total & Q & M & All \% \\
    \midrule
    robot\_fault\_prediction & 17 & 7/10 & 6/7 & 76.5 \\
    reverse\_logistics & 28 & 13/20 & 8/8 & 75.0 \\
    cybermarket\_pattern & 30 & 12/20 & 9/10 & 70.0 \\
    solar\_panel & 30 & 13/20 & 8/10 & 70.0 \\
    crypto\_exchange & 30 & 12/20 & 8/10 & 66.7 \\
    sports\_events & 29 & 11/20 & 8/9 & 65.5 \\
    exchange\_traded\_funds & 28 & 13/19 & 4/9 & 60.7 \\
    fake\_account & 30 & 12/24 & 6/6 & 60.0 \\
    hulushows & 30 & 10/20 & 8/10 & 60.0 \\
    museum\_artifact & 30 & 7/20 & 9/10 & 53.3 \\
    virtual\_idol & 28 & 7/19 & 7/9 & 50.0 \\
    households & 29 & 6/21 & 8/8 & 48.3 \\
    polar\_equipment & 30 & 7/20 & 7/10 & 46.7 \\
    planets\_data & 29 & 10/19 & 3/10 & 44.8 \\
    cross\_border & 29 & 7/19 & 5/10 & 41.4 \\
    insider\_trading & 30 & 9/21 & 3/9 & 40.0 \\
    archeology\_scan & 13 & 3/10 & 2/3 & 38.5 \\
    disaster\_relief & 15 & 4/12 & 1/3 & 33.3 \\
    cold\_chain\_pharma\_compliance & 28 & 5/18 & 4/10 & 32.1 \\
    labor\_certification\_applications & 29 & 6/19 & 3/10 & 31.0 \\
    organ\_transplant & 32 & 2/19 & 6/13 & 25.0 \\
    mental\_health & 26 & 2/20 & 3/6 & 19.2 \\
    \midrule
    \textbf{Total} & \textbf{600} & \textbf{178/410} & \textbf{126/190} & \textbf{50.7} \\
    \bottomrule
  \end{tabular}
  \caption{LiveSQLBench per-database accuracy on the
    600-instance complete run.
    Columns: total instances, Query category
    (correct/total), Modification category (correct/total),
    overall accuracy. Seven databases score above 60\% and six
    below 40\%; the bottom tier is dominated by domain ambiguity
    in the user knowledge base (\texttt{mental\_health},
    \texttt{organ\_transplant},
    \texttt{labor\_certification\_applications}), while the top
    tier is well-specified operational data with clean joins
    (\texttt{robot\_fault\_prediction}, \texttt{reverse\_logistics},
    \texttt{cybermarket\_pattern}).}
  \label{tab:appE-livesqlbench}
\end{table}

\begin{table}[H]
  \centering
  \small
  \begin{tabular}{l r r r}
    \toprule
    Database & Total & Correct & Accuracy (\%) \\
    \midrule
    esophageal & 8 & 7 & 87.5 \\
    california\_schools & 24 & 18 & 75.0 \\
    toxicology & 64 & 46 & 71.9 \\
    card\_games & 51 & 35 & 68.6 \\
    financial & 44 & 30 & 68.2 \\
    debit\_card\_specializing & 37 & 25 & 67.6 \\
    global\_atlas & 3 & 2 & 66.7 \\
    thrombosis\_prediction & 32 & 20 & 62.5 \\
    formula\_1 & 44 & 27 & 61.4 \\
    superhero & 40 & 24 & 60.0 \\
    codebase\_community & 12 & 7 & 58.3 \\
    european\_football\_2 & 69 & 40 & 58.0 \\
    erolp & 28 & 16 & 57.1 \\
    student\_club & 42 & 23 & 54.8 \\
    spotify & 2 & 1 & 50.0 \\
    \midrule
    \textbf{Total} & \textbf{500} & \textbf{321} & \textbf{64.2} \\
    \bottomrule
  \end{tabular}
  \caption{BIRD-Critic per-database accuracy on the
    500-instance complete run. The Management category is the
    cleanest tier: the agent benefits substantially from
    declaring the expected result shape before generation and from
    the identifier extractor that surfaces backtick-quoted names
    verbatim from the question text. Personalization is the
    hardest because gold often goes beyond what the question
    explicitly enumerates (e.g.\ adds derived columns, returns
    nested JSON instead of a flat result set), and the agent has
    no signal to predict gold's exact shape without seeing the
    test code.}
  \label{tab:appE-birdcritic}
\end{table}

\begin{table}[H]
  \centering
  \small
  \begin{tabular}{l r r r}
    \toprule
    Database & Total & Correct & All \% \\
    \midrule
    cybermarket\_pattern & 30 & 26 & 86.7 \\
    reverse\_logistics & 28 & 24 & 85.7 \\
    fake\_account & 30 & 25 & 83.3 \\
    insider\_trading & 30 & 21 & 70.0 \\
    crypto\_exchange & 30 & 20 & 66.7 \\
    solar\_panel & 30 & 20 & 66.7 \\
    exchange\_traded\_funds & 28 & 17 & 60.7 \\
    disaster\_relief & 15 & 9 & 60.0 \\
    planets\_data & 29 & 17 & 58.6 \\
    households & 29 & 16 & 55.2 \\
    sports\_events & 29 & 16 & 55.2 \\
    hulushows & 30 & 16 & 53.3 \\
    virtual\_idol & 28 & 14 & 50.0 \\
    labor\_certification\_applications & 29 & 14 & 48.3 \\
    robot\_fault\_prediction & 17 & 8 & 47.1 \\
    polar\_equipment & 30 & 14 & 46.7 \\
    cold\_chain\_pharma\_compliance & 28 & 13 & 46.4 \\
    archeology\_scan & 13 & 5 & 38.5 \\
    cross\_border & 29 & 11 & 37.9 \\
    organ\_transplant & 32 & 11 & 34.4 \\
    museum\_artifact & 30 & 10 & 33.3 \\
    mental\_health & 26 & 7 & 26.9 \\
    \midrule
    \textbf{Total} & \textbf{600} & \textbf{334} & \textbf{55.7} \\
    \bottomrule
  \end{tabular}
  \caption{BIRD-Interact per-database success rate on the
    600-instance run. Phase, category, and difficulty aggregates
    are in Table~\ref{tab:appE-breakdown}; results by follow-up
    type are in Table~\ref{tab:appE-followup}.}
  \label{tab:appE-birdinteract}
\end{table}

\begin{table}[H]
  \centering
  \small
  \begin{tabular}{l r r r r r}
    \toprule
    Database & N & Correct & EX & soft-F1 & VES \\
    \midrule
    superhero & 129 & 117 & 90.7 & 92.9 & 83.9 \\
    student\_club & 158 & 143 & 90.5 & 92.2 & 84.5 \\
    debit\_card\_specializing & 64 & 54 & 84.4 & 88.4 & 78.3 \\
    codebase\_community & 186 & 153 & 82.3 & 85.0 & 76.8 \\
    european\_football\_2 & 129 & 101 & 78.3 & 81.0 & 72.8 \\
    toxicology & 145 & 111 & 76.6 & 75.7 & 70.4 \\
    card\_games & 191 & 145 & 75.9 & 75.8 & 70.8 \\
    formula\_1 & 174 & 130 & 74.7 & 74.3 & 68.8 \\
    california\_schools & 89 & 64 & 71.9 & 77.5 & 66.4 \\
    thrombosis\_prediction & 163 & 107 & 65.6 & 67.4 & 58.4 \\
    financial & 106 & 67 & 63.2 & 69.2 & 59.0 \\
    \midrule
    \textbf{Total} & \textbf{1{,}534} & \textbf{1{,}192} & \textbf{77.7} & \textbf{79.2} & \textbf{71.9} \\
    \bottomrule
  \end{tabular}
  \caption{BIRD-Dev per-database results, sorted by execution
    accuracy.}
  \label{tab:appE-birddev-perdb}
\end{table}

\begin{table}[H]
  \centering
  \small
  \begin{tabular}{l r r r r}
    \toprule
    Benchmark & DBs & Fully solved & Partially solved & Unsolved \\
    \midrule
    Spider2-Lite & 88 & 36 & 37 & 15 \\
    Spider2-Snow & 152 & 63 & 64 & 25 \\
    Spider2-DBT & 64 & 24 & --- & 40 \\
    \bottomrule
  \end{tabular}
  \caption{Distribution of per-database outcomes on the Spider2
    family: databases where every question is answered correctly,
    where some are, and where none are. Spider2-DBT instances are
    single whole-project completions, so no partial bucket
    applies.}
  \label{tab:appE-spider2dist}
\end{table}

\section{Error Analysis}
\label{sec:appC}
\label{app:errors}

This appendix expands the error analysis of
Section~\ref{sec:eval-errors}. The finding that organizes it is
that the remaining headroom is semantic rather than syntactic: on
every benchmark almost every failed instance executes cleanly and
returns a wrong answer. Almost all of them are semantic, and we
sort them into three recurring classes, each pointing to a
different remedy. Reasoning failures, where the
query answers a subtly different question than the one asked, are
the largest class on every benchmark and are bound to model
capability. Output-convention failures, where the right quantity is
computed but presented in the wrong shape, are the most addressable
class and respond to refinements of the standing instructions.
Grounding failures, where the agent cannot resolve an exact
identifier, appear on the benchmarks with object-modification tasks
and recur within a database, the recurrence structure that memory
targets (Section~\ref{sec:eval-memory}). Execution failures, where
the SQL does not run at all, are rare, under three percent of
failures in aggregate.

We classify every failure of each benchmark's reported run by
comparing the agent's output to gold, directly from the
per-instance results: 342 failures on BIRD-Dev, 296 on
LiveSQLBench, 266 on BIRD-Interact, 179 on BIRD-Critic, 98 on
Spider2-Lite, 167 on Spider2-Snow, and 40 on Spider2-DBT. Where the
results record executed result tables, on BIRD-Dev, LiveSQLBench
query tasks, and the Spider2 family, we compare predicted and gold
result shapes and values; where they record only pass or fail
against hidden tests, on BIRD-Critic and the modification tasks, we
compare the predicted and gold SQL. Two limits follow. Output
convention is detected only where the result shape is observable,
so on the SQL-only failures its share is a lower bound; and a wrong
literal value, which is a grounding error in spirit, is
indistinguishable from wrong logic once it reaches the result, so
on the pure query benchmarks it falls under reasoning.

\subsection{Failure classes across benchmarks}

\begin{figure}[t]
  \centering
  \includegraphics[width=\textwidth]{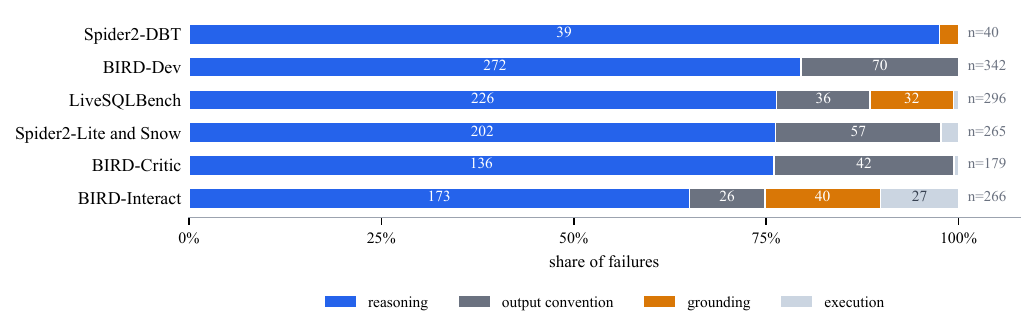}
  \caption{Composition of failures per benchmark, aggregated over
    each benchmark's task categories and ordered by the share of
    reasoning failures. Segment labels are failure counts.}
  \label{fig:appC-errors}
\end{figure}

Figure~\ref{fig:appC-errors} reports the class composition of
failures per benchmark.

Reasoning failures dominate every benchmark, from roughly three
quarters on most to two thirds of BIRD-Interact's. The class covers
three recurring shapes. Selection errors choose the wrong rows
through a wrong filter, join path, or entity, and account for the
largest buckets of Table~\ref{tab:appC-livesql}. Computation errors
feed the right rows into a wrong formula or aggregate at the wrong
granularity, and drive the family-wide signature of same-shape
results with wrong values. Construct errors misuse a SQL idiom,
such as an inner join where the question implies an outer one, ties
dropped at a top-N boundary, or a defensive NULL filter that
removes valid rows, and are among the structurally identifiable
patterns of Table~\ref{tab:appC-critic}. All three execute cleanly and
concentrate where the question under-specifies its intent; none
respond to instruction refinements, marking the capability frontier
of the model rather than a fixable gap in the system.

Output-convention failures are the largest addressable class. Extra
or missing output columns are the main form, with row order making
up the rest. The computed quantity is right and
only its presentation diverges from what the gold answer admits,
which is why these patterns respond to standing-instruction
refinements where reasoning failures do not. Because convention is
counted only where the result shape is observable, its share on the
SQL-only benchmarks understates the true total.

Grounding failures concentrate on the benchmarks with
object-modification tasks. Most of LiveSQLBench's and
BIRD-Interact's grounding failures trace to table, view, or column
names the agent paraphrased, so a probe written against the gold
name cannot find them. These recur within a database, the same
identifiers returning question after question, which is the
recurrence structure that memory targets
(Section~\ref{sec:eval-memory}).

\subsection{Per-benchmark patterns}

Tables \ref{tab:appC-livesql} and \ref{tab:appC-critic} give the
per-pattern breakdown for the two benchmarks whose results expose
the most detail: LiveSQLBench, which records executed result tables
for its query tasks, and BIRD-Critic, whose pass or fail outcomes
we read through the predicted and gold SQL.

\begin{table}[H]
  \centering
  \small
  \begin{tabular}{l r >{\raggedright\arraybackslash}p{7.6cm}}
    \toprule
    Pattern & Count & Diagnosis \\
    \midrule
    wrong rows returned        & 105 & wrong filter, join path, or entity: the returned rows barely overlap gold \\
    row count moderately off   &  58 & wrong filter or join cardinality, within a factor of five \\
    row count far off          &  22 & major logic error, off by more than a factor of five \\
    extra output columns       &  16 & display columns the gold answer does not admit \\
    missing output columns     &  14 & projection narrowed past the requested columns \\
    right rows, wrong values   &   7 & correct rows but wrong computation or formula \\
    wrong row order            &   6 & correct rows in the wrong order \\
    empty result               &   4 & the query returned no rows \\
    \bottomrule
  \end{tabular}
  \caption{LiveSQLBench query failures (232), by comparison of the
    executed result with gold. The 64 modification failures are
    test-case-graded and split into grounding (32, paraphrased
    identifiers), reasoning (30, logic errors), and execution (2).}
  \label{tab:appC-livesql}
\end{table}

\begin{table}[H]
  \centering
  \small
  \begin{tabular}{l r >{\raggedright\arraybackslash}p{7.6cm}}
    \toprule
    Pattern & Count & Diagnosis \\
    \midrule
    extra output columns    & 23 & output decorated with the question's free-text dimensions \\
    missing output columns  & 19 & projection narrowed past the columns the user named \\
    unrequested NULL filter & 23 & defensive \texttt{IS NOT NULL} that drops valid rows \\
    dropped ties            &  8 & \texttt{ROW\_NUMBER() = 1} or \texttt{LIMIT 1} where gold keeps ties \\
    overcomplicated SQL     &  5 & recursive CTEs or manual date arithmetic where a simpler idiom suffices \\
    wrong join type         &  2 & \texttt{INNER} where the question implies \texttt{LEFT}, or the reverse \\
    \bottomrule
  \end{tabular}
  \caption{BIRD-Critic failures with a structurally identifiable
    pattern in the SQL (80 of 179). The remaining 98 are value or
    logic errors not separable from the SQL text alone, and one is
    an execution error. Extra or missing output columns, the
    convention class, are the largest identifiable patterns and
    recur across the query, personalization, and management
    categories alike.}
  \label{tab:appC-critic}
\end{table}

For the Spider2 family the same result-shape comparison applies
directly. Of Spider2-Lite and Spider2-Snow's 265 query failures,
202 return a result of gold's shape with wrong values or row counts,
57 differ in column count, and 6 fail to execute. The 40
Spider2-DBT project failures, taken at the first mismatching table
of the built database, are almost all content errors: 39 build the
required tables but with wrong contents, and 1 leaves a required
table uncreated so a downstream reference errors. Because dbt grades
by whole-table equality with no result tuples to inspect,
column-count divergences cannot be separated from content errors
here, so the convention share folds into the first bucket and the
lower-bound caveat above applies most strongly to this benchmark.
Across the family wrong values
and row counts dominate, extra or missing columns are the main
addressable slice, and outright execution errors stay near two
percent, consistent with Section~\ref{sec:eval-errors}.

\subsection{Interaction failure behaviour}

BIRD-Interact adds a behavioural failure dimension that the
single-shot benchmarks lack. Its 266 failures split into 215 in
phase 1 and 51 in phase 2.

The dominant behaviour is hypothesis lock-in: 215 of the 266
failures (81\%) are phase-1 trajectories that exhausted the 15-turn
budget, the agent re-submitting near-identical SQL against the
protocol's minimal feedback instead of stepping back to ask a
clarifying question. The recurring modes are inferring a composite
formula without asking, chasing sort order when the real divergence
was a formula input, and phase-1 views whose paraphrased column
names break the gold phase-2 SQL. These modes map onto the same
classes as the single-shot benchmarks: composite-formula inference
is a reasoning failure, sort-order chasing an output-convention
failure, and paraphrased identifiers a grounding failure.
Classifying the final submissions accordingly gives the
BIRD-Interact bar of Figure~\ref{fig:appC-errors}: 173 reasoning, 40
grounding from paraphrased identifiers on management tasks, 26
output-convention, and 27 execution errors from malformed
submissions at the turn cap. Worked traces of the recurring modes
appear in Appendix~\ref{sec:appD}, together with the interaction
policies they motivated.
\section{Case Studies}
\label{sec:appD}

The premise of this work is that an agent grounded in execution
can answer reliably because it can verify its work against the
database. BIRD-Interact is the sharpest test of that premise, and
the benchmark where DIA's margin over prior work is largest
(Table~\ref{tab:main}), because it is the one setting where part
of the ground truth does not live in the database: the user's
intended formula and the canonical names of requested artifacts
are known only to the user. Execution grounding answers
everything the data can answer; dialogue must cover the rest, and
knowing the boundary between the two is the skill the benchmark
rewards. It is also the benchmark that most resembles our
production deployment, where domain experts pose under-specified
questions conversationally and clarification is part of normal
operation (Appendix~\ref{app:deploy}). This appendix shows that boundary in worked traces: one
released-run pass where grounding and a single targeted question
divide the work correctly, and two failure modes, a stuck-loop and
a Phase-2 cascade, that we observed during development and that
motivated the interaction policies described at the end. We pair
each failure trace with the released-run outcome of that same
instance once the policies are in place.

\subsection{The BIRD-Interact protocol}

Each BIRD-Interact instance is a multi-turn conversation between
the \emph{Query Generator} and the benchmark's LLM-driven user
simulator. Three properties make it
qualitatively different from the other six benchmarks in our evaluation set:

\begin{itemize}
  \item Two phases per instance. Phase~1 is the
    primary question (\texttt{Q}~or~\texttt{M}); Phase~2 is a
    follow-up that builds on the agent's Phase~1 answer. The agent
    must SUBMIT a SQL for Phase~1; if it passes, Phase~2 issues a
    new prompt that references Phase~1's result shape (e.g.\ ``filter
    the result you just produced to rows where $\ldots$'').
  \item An ASK or SUBMIT protocol with a fixed turn cap per
    phase (Appendix~\ref{app:impl}). Every turn the agent either emits
    \texttt{ASK: <question>} to request clarification or
    \texttt{SUBMIT: <sql>} to attempt an answer. ASKs are routed
    to the user simulator (an LLM playing the role of a non-SQL
    domain expert); SUBMITs are evaluated, and an
    INCORRECT verdict returns a structural delta (row count,
    column count) without disclosing gold values. This is the
    benchmark's information-asymmetric protocol, matching what
    real users could actually provide.
  \item Follow-ups bind to Phase-1 artifacts. For Management
    tasks, Phase~2 references the object the agent created in
    Phase~1, including its column names. This surfaces a cascade
    failure mode unique to BIRD-Interact: if Phase~1 paraphrased
    a column name, the Phase~2 evaluation cannot see it
    (see the Phase-2 cascade trace).
\end{itemize}

The combination produces a benchmark where 81\% of failures burn
the full turn cap on a single phase (the agent keeps re-submitting
near-identical SQL against the same minimal feedback), and where
the decisive skill is not which SQL to emit
but when to stop guessing and ask.

\begin{figure}[t]
  \centering
  \includegraphics[width=0.72\textwidth]{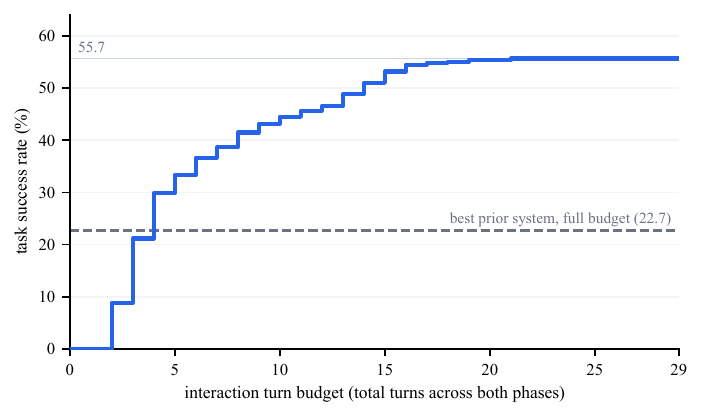}
  \caption{Interaction-time scaling on BIRD-Interact: the
    fraction of the 600 instances whose passing submission landed
    within the first $k$ total turns across both phases, computed
    from the single full-budget run. The dashed line is the final
    score of the best prior system at its full budget.}
  \label{fig:appD-turns}
\end{figure}

Figure~\ref{fig:appD-turns} gives a single-run view of what the
benchmark's authors term interaction-time scaling
\citep{birdinteract}, the ability to convert multi-turn
interaction into end performance: rather than re-running the
benchmark at each budget, it reports the turn at which each
passing submission landed under the full budget. The curve
is cumulative, so what is informative is its shape: DIA exceeds
the final score of the best prior system within its first four
turns, is still climbing steeply late in the budget (turns eleven
through fifteen add another nine points), and flattens only past
twenty total turns, so the cap leaves little on the table. We do
not evaluate the benchmark's stronger criterion (the
interaction-time scaling law), which compares interactive
performance against idealized single-turn performance on the
disambiguated task. Successful instances are efficient, finishing
in a median of four turns, while failed instances are the ones
pinned at the cap. Under-specification also has a measurable
dialogue cost even in success: passing high-level instances need
a median of six turns against four for low-level ones, the price
of resolving a composite metric through conversation.

\subsection{A passing trace}

The passing trace shows the division of labour working: one
targeted ASK for the formula that execution cannot reveal,
workspace probes for everything it can, and a first-submission
pass on both phases.

\begin{tracebox}{Passing trace (archeology\_scan instance 1)}
TURN 1
USER: For each scan site, report the average ambient temperature,
humidity, and illumination, along with the Environmental
Sensitivity Index (ESI). Order by ESI descending.

AGENT: ASK: The KB defines ESI in terms of three sub-metrics
(ambient_temp_norm, humidity_norm, illumination_norm). Could you
confirm whether ESI is the simple mean of the three normalized
values, or a weighted combination?

USER_SIM: labeled("ESI"). Each row of the scans table has an
ambient_cond JSON object with Ambic_Temp, Ambic_Humidity, and
Ambic_Light keys. ESI is the simple mean of the three normalized
values, after each value is min-max normalized over the entire
table. Use COALESCE(<expr>::real, <default>) with the defaults
listed in the KB entry for ESI when a key is missing.

TURN 2
AGENT (workspace): writes and runs a probe script before
submitting:

  keys = query_db("SELECT DISTINCT jsonb_object_keys(ambient_cond)
                   FROM scans")
  # -> Ambic_Temp, Ambic_Humidity, Ambic_Light

  missing = query_db("SELECT COUNT(*) FROM scans
                      WHERE NOT ambient_cond ? 'Ambic_Temp'")
  # -> some rows lack the key: the COALESCE defaults matter

AGENT: SUBMIT:
WITH normed AS (
  SELECT s.arcref AS site_id,
         COALESCE((s.ambient_cond->>'Ambic_Temp')::real, 20.0) AS t,
         COALESCE((s.ambient_cond->>'Ambic_Humidity')::real, 50.0) AS h,
         COALESCE((s.ambient_cond->>'Ambic_Light')::real, 300.0) AS l
  FROM scans s
),
bounds AS (
  SELECT MIN(t) AS t_min, MAX(t) AS t_max,
         MIN(h) AS h_min, MAX(h) AS h_max,
         MIN(l) AS l_min, MAX(l) AS l_max
  FROM normed
)
SELECT n.site_id,
       AVG(n.t) AS avg_temp,
       AVG(n.h) AS avg_humidity,
       AVG(n.l) AS avg_illumination,
       AVG( (n.t - b.t_min) / NULLIF(b.t_max - b.t_min, 0)
          + (n.h - b.h_min) / NULLIF(b.h_max - b.h_min, 0)
          + (n.l - b.l_min) / NULLIF(b.l_max - b.l_min, 0) ) / 3.0 AS esi
FROM normed n CROSS JOIN bounds b
GROUP BY n.site_id
ORDER BY esi DESC;

EVALUATOR: PASS (rows=147, columns=5, match within 1 ULP)

PHASE 2 prompt: Restrict the result to the top 10 most-sensitive
sites and add a classification column ('high'/'medium'/'low')
based on NTILE(3) over ESI.

AGENT: SUBMIT:
WITH p1 AS (
  -- accepted phase-1 SQL, carried forward as the base
  ... [as above] ...
)
SELECT site_id, avg_temp, avg_humidity, avg_illumination, esi,
       CASE NTILE(3) OVER (ORDER BY esi DESC)
         WHEN 1 THEN 'high' WHEN 2 THEN 'medium' ELSE 'low' END
         AS sensitivity_class
FROM p1
ORDER BY esi DESC
LIMIT 10;

EVALUATOR: PASS. Final verdict: PASS (both phases).
\end{tracebox}

\noindent Three properties of this trace are diagnostic of when
the agent succeeds: (i) it divides the work correctly between
grounding and dialogue, probing the database for what execution
can answer (key spellings, missing-key rates) and ASKing only for
what it cannot (the formula definition); (ii) it uses the
KB-listed COALESCE defaults verbatim (the stuck-loop trace
shows the failure mode this avoids); and (iii) Phase~2 wraps the
accepted Phase~1 SQL in a CTE rather than rewriting from scratch
(DIA carries the accepted Phase-1 SQL into the follow-up context,
making this the natural default).

\subsection{A stuck-loop failure}

This trace, recorded during development before the
clarification policy below was in place, shows what was then the
most common BIRD-Interact failure mode: the agent infers a
composite formula from KB names without asking, then burns the
entire turn budget with cosmetic edits (cast types, JSON-key
spellings) instead of stepping back to ASK.

\begin{tracebox}{Stuck-loop trace (archeology\_scan instance 5, development run): turn budget exhausted with 0 ASKs}
KB CONTEXT: DPQ = 0.4*SQS + 0.4*MFS + 0.3*SCE_mod + 0.2*RAR,
where each sub-term is itself a multi-term expression over
JSON keys masked from the agent.

TURN 1
USER: For each scan location, compute a Data Provenance Quality
index (DPQ).

AGENT: SUBMIT: -- guessed composite from KB term names
SELECT zoneref,
       0.4*sqs + 0.4*mfs + 0.3*sce + 0.2*rar AS dpq
FROM (
  SELECT zoneref,
         AVG((meta->>'survey_quality_score')::real)    AS sqs,
         AVG((meta->>'metadata_fidelity_score')::real) AS mfs,
         AVG((meta->>'site_completeness_eval')::real)  AS sce,
         AVG((meta->>'reference_alignment')::real)     AS rar
  FROM scans GROUP BY zoneref
) t
ORDER BY dpq DESC;

EVALUATOR: INCORRECT. (Row and column counts match gold;
  values differ. No gold values disclosed.)

TURN 2 ... TURN 14: agent re-SUBMITs 14 near-identical variants:
swaps `::real` for `::numeric`, varies the JSON keys
(`survey_quality_score` -> `sqs_raw` -> `surveyqualscore`),
wraps in additional CTEs, adds ORDER BY ties. Every SUBMIT
returns the same verdict, with no gold values disclosed.

TURN 15 (final turn)
AGENT: SUBMIT: -- final cosmetic variant, still wrong
... [as before, with one cast rewritten] ...

EVALUATOR: INCORRECT. Final verdict: FAIL (Phase-1 not passed,
Phase-2 not attempted).
\end{tracebox}

\noindent The agent never ASKs to clarify either (a) what the
masked sub-terms actually compute (the KB lists their names but
not the formulas they expand to) or (b) whether the composite is
computed per-row before aggregation or after. Both are required to
reach gold's roughly 5{,}000-character composite. This is a
boundary failure of execution grounding: the formula's semantics
live with the user, not in the data, so no amount of probing or
re-execution can verify a guessed composite. The agent's mistake
is treating a dialogue problem as a grounding problem. Standing
guidance alone does not prevent this mode: nothing behavioural
forces the agent to back off from a losing SUBMIT to an ASK,
which is what motivated the clarification-question policy. With
that policy active in the released run, pure stuck loops are
nearly eliminated: only two of the 215 Phase-1 failures issue as
few as one clarifying question and none issue zero, and this same
instance now issues several ASKs before exhausting its budget. The
trace shows the failure mode the policy targets, not residual
released behaviour.

\subsection{A Phase-2 cascade}

This trace, also from the development run, shows a class of failure
unique to BIRD-Interact Management: the agent's Phase-1 view shape
is locally correct but missing a column that the Phase-2 evaluation
will then need to reference.

\begin{tracebox}{Phase-2 cascade trace (archeology\_scan M\_3, development run)}
PHASE 1
USER: Create a view `view_premium_quality_scans` containing scan
ID, project, site, timestamp, resolution, density, coverage,
overlap, and noise, restricted to scans whose DPQ is above the
75th percentile.

AGENT: SUBMIT:
CREATE OR REPLACE VIEW view_premium_quality_scans AS
WITH q AS (... DPQ computation ...)
SELECT s.facetregistry AS scan_id,
       s.project,
       s.facetregistry AS site_code,    -- paraphrased "site"
       s.captured_at  AS timestamp,
       s.resolution_m AS resolution,
       ...
FROM scans s JOIN q USING (zoneref)
WHERE q.dpq > q.dpq_p75;

EVALUATOR: PASS (Phase 1 view created; rows match gold count).

PHASE 2
USER: For each project in `view_premium_quality_scans`, report
the count of premium scans and the average noise level.

AGENT: SUBMIT:
SELECT project, COUNT(*) AS n, AVG(noise) AS avg_noise
FROM view_premium_quality_scans
GROUP BY project;

EVALUATOR: the Phase-2 evaluation references the view's
underlying column names:
  SELECT zoneref, COUNT(*), AVG(noise)
  FROM view_premium_quality_scans GROUP BY zoneref;
... which raises UndefinedColumn: column
view_premium_quality_scans.zoneref does not exist.
(The agent aliased site to site_code.)

Verdict: FAIL (Phase-2 cascade due to Phase-1 column-name drift).
\end{tracebox}

\noindent The agent's Phase-1 view is functionally correct: the
row set matches gold and the projection covers the columns the
question enumerated. It fails not because of its own Phase-2
SQL but because the follow-up's expected answer is keyed to the
underlying column names: the Phase-2 evaluation references
\texttt{zoneref} (the table's actual column name) while the agent
paraphrased it to \texttt{site\_code}. This is the other boundary
failure of execution grounding: execution can validate that the
view runs and its rows match, but it cannot reveal the canonical
names a follow-up will expect, because naming is a matter of user
intent rather than data. DIA's standing
instructions therefore treat a Management projection list as not
a renaming contract: when the question names a column
conversationally (``site''), the agent projects the underlying
column under its original identifier rather than inventing a
conversational alias. With this policy in place, the same instance
passes both phases in the released run: the Phase-1 view exposes
\texttt{zoneref} under its own name, so the Phase-2 follow-up
resolves.

\subsection{Interaction policies}

These failure modes shaped how the \emph{Query Generator}
behaves in multi-turn protocols. Four mechanisms carry most of
the weight.

\paragraph{Forced clarification.}
After three consecutive INCORRECT verdicts with the same
structural signature, DIA requires its next action to be an ASK
rather than another SUBMIT. Stuck loops are the largest
behavioural failure mode, and this policy cuts them directly. The
streak resets on any ASK or on a SUBMIT whose verdict differs.
Ablating this policy across the full 600-instance run drops
accuracy from 55.7\% to 14.5\%, a 41.2-point fall, confirming that
stuck loops, not a limit on the model's SQL ability, are what the
policy is protecting against.

\paragraph{Resubmission control.}
DIA does not resubmit near-identical SQL: a candidate SUBMIT that
matches one of its recent attempts is rejected, and the agent
must either ASK or change the structural approach. This closes
the repeated-resubmission pattern within stuck-loop traces.

\paragraph{Context carryover.}
The accepted Phase-1 SQL is placed at the head of the Phase-2
context, so the natural default is to apply the requested
follow-up edit to the Phase-1 base, which is the protocol's
intent, rather than rewriting from scratch and silently changing
the Phase-1 filter or projection.

\paragraph{Short-term and long-term memory.}
Memory extends the same boundary, and in deployment we see this
in practice. Within a conversation, clarifications act as
short-term memory: once the user states a formula, every later
turn builds on it. Across conversations, those answers become
long-term memory: the same domain experts return with the same
vocabulary, and a formula or naming convention they have already
explained is not asked for again. Our treatment of memory
(Section~\ref{sec:eval-memory}) concerns recurrence within a
database rather than within a user.

Together these mechanisms encode the boundary the traces
illustrate: the agent grounds everything the database can answer
by execution and spends dialogue only on what it cannot. That
division of labour, rather than any difference in SQL fluency, is
where the margin on this benchmark comes from
(Table~\ref{tab:main}).

\section{Standing Instructions}
\label{sec:appA}

The standing instructions of the \emph{Query Generator} share
one architecture across benchmarks. Every seed file combines
three ingredients: an output-contract discipline, under which the
agent declares the expected shape of its answer and verifies the
executed result against it; reference material and pitfalls for
the SQL dialect; and guidance for the task format. The
per-question prompt itself is thin: it lays out the workspace,
states the question and its task metadata, and points the agent
at the seed. Table~\ref{tab:appA-seeds} summarizes how each
benchmark instantiates this architecture.

\begin{table}[H]
  \centering
  \small
  \begin{tabular}{l >{\raggedright\arraybackslash}p{9.2cm}}
    \toprule
    Benchmark & Task-specific content of its seed \\
    \midrule
    BIRD-Dev      & projection discipline; verification of literal values and transforms against the live database; memory usage \\
    BIRD-Critic   & a five-step debugging workflow; patterns for the query, personalization, and management categories \\
    LiveSQLBench  & the same workflow adapted to generation; data-modification guidance; external-knowledge usage rules \\
    BIRD-Interact & conversational guidance: when to ask versus when to submit, what the simulated user can answer, and management-task fidelity \\
    Spider2-Lite  & a compact output contract and semantic rules \\
    Spider2-Snow  & the same contract instantiated for Snowflake \\
    Spider2-DBT   & a project-completion workflow: analyze the dbt project, build, and verify the produced database \\
    \bottomrule
  \end{tabular}
  \caption{How each benchmark instantiates the shared
    architecture: the task-specific content of its seed, beyond
    the output-contract discipline and the dialect reference.}
  \label{tab:appA-seeds}
\end{table}

\subsection{The workflow skeleton}

The fullest form of the contract discipline is a five-step
workflow used by the debugging, modification, and conversational
benchmarks: plan with an explicit contract, diagnose by
execution, build, validate against the contract, save. We
reproduce a condensed BIRD-Critic instantiation.

\begin{tracebox}{Workflow skeleton as instantiated for BIRD-Critic (STEP 1 to STEP 5)}
## Per-Question Workflow

### STEP 1 - PLAN
- Read the user's question and any buggy SQL carefully.
- Identify the category (Query / Personalization / Management).
- Define an explicit OUTPUT_CONTRACT:
    cols:    [c1, c2, ...]
    rows:    ~N (one per <entity>)
    order:   <unordered | by X asc/desc>
    filters: [every constraint expressed in the question]
- Filters checklist (CRITICAL): walk every adjective, prepositional
  phrase, "only/excluding/ignoring", and conjunction in the prose.
  Each becomes a row in `filters: [...]`.

### STEP 2 - DIAGNOSE
- Inspect the schema (PRAGMA table_info / information_schema).
- Execute the buggy SQL (if any) and observe its actual output.
- Compare actual output to OUTPUT_CONTRACT. What's wrong?

### STEP 3 - BUILD
- Projection: entity-key -> metric -> drop redundant display cols
  -> drop helper window cols from outer SELECT.
- Minimal-diff: smallest edit to buggy SQL that fixes the bug.
- Apply every filter in the STEP 1 checklist.

### STEP 4 - VALIDATE
- Run pred via query_db; observe the result set.
- For every item in `filters`, point at the clause that implements it.
- Check column count, row count plausibility, JOIN cardinality,
  first-row sanity, and absence of spurious NULL filters.

### STEP 5 - SAVE
- save_result(corrected_sql, result_rows, tables_used)
- For Management tasks save the DDL/DML statements themselves
  (not the verification SELECT).
- cleanup() to reset the workspace for the next question.
\end{tracebox}

The contract is declared before any SQL is run, so STEP 4 can
fail an answer on shape grounds using nothing but the agent's own
plan. This is the self-verification loop of the main paper in
operational form, and it targets the largest addressable failure
class of Appendix~\ref{app:errors}: extra display columns,
missing filters, and helper columns leaking into the outer
projection.

\subsection{A worked debugging example}

The trace below is an illustrative reconstruction, composed to show
the five-step contract discipline end to end rather than transcribed
from a single released instance. It walks the workflow through a
representative BIRD-Critic debugging question over the
\texttt{financial} database: the agent receives the user's prose and
a buggy SQL fragment, diagnoses the bug by executing it, and repairs
it with a minimal edit.

\begin{tracebox}{Worked trace (illustrative; BIRD-Critic, financial database)}
QUESTION: For each account, what is the latest transaction date
where the amount exceeded the account's average transaction
amount? List account_id and the transaction date.

BUGGY SQL (issued by the user):
  SELECT a.account_id, MAX(t.date)
  FROM account a
  JOIN trans t ON a.account_id = t.account_id
  WHERE t.amount > AVG(t.amount)
  GROUP BY a.account_id;

STEP 1 PLAN: the agent declares its contract.
  OUTPUT_CONTRACT:
    cols:    [account_id, latest_date]
    rows:    ~one per account (~4500)
    order:   none specified
    filters: [transactions whose amount > account-level mean]

STEP 2 DIAGNOSE: the agent executes the buggy SQL and observes
  SQLite error: misuse of aggregate function AVG()
The aggregate sits inside a row-level WHERE, which is invalid:
the per-account mean must be computed per group.

STEP 3 BUILD: minimal repair with a window function.
  WITH t_with_avg AS (
    SELECT t.account_id, t.date, t.amount,
           AVG(t.amount) OVER (PARTITION BY t.account_id) AS acct_avg
    FROM trans t
  )
  SELECT account_id, MAX(date) AS latest_date
  FROM t_with_avg
  WHERE amount > acct_avg
  GROUP BY account_id;

STEP 4 VALIDATE: the agent runs the corrected SQL via query_db.
  4500 rows, 2 columns
Contract check: column count matches (2 = 2); one row per
account; the filter "amount above account-level mean" is
implemented by WHERE amount > acct_avg; no spurious
IS NOT NULL filter was added.

STEP 5 SAVE: save_result(corrected_sql, result_rows,
  tables_used=[trans, account]); cleanup().
\end{tracebox}

The repair is the minimal edit the seed asks for: the structure
of the buggy SQL is preserved, and only the invalid aggregate
placement changes. The contract declared in STEP 1 is what makes
STEP 4 a real check rather than a formality: every entry in it is
verified against the executed result before the answer is saved.
\section{Memory: Store and Contents}
\label{app:memory}

This appendix gives concrete form to the memory introduced in
Section~\ref{sec:memory}: how the experience store is organized, how its
entries are written, and what kinds of knowledge it captures during a run.
The view is observational, drawn from the BIRD-Dev run.

\subsection{The three tiers}

Memory is held as files in the workspace and organized in three tiers.

\paragraph{Retrieved examples.} The pool for this tier is the BIRD training
split: its gold question-and-SQL pairs together with per-table column-meaning
notes, embedded once offline into a fixed similarity index that is reused
unchanged across runs. The split is disjoint from the BIRD-Dev evaluation set,
so retrieval introduces no test leakage. For each question, the few most similar
pairs above a fixed similarity threshold (Appendix~\ref{app:impl}) are surfaced
and staged for the agent to consult. These are concrete examples rather than
distilled rules: the agent reads them as idioms and re-derives any abstraction at
the point of use, confirming column names and values against the current database
before relying on them.

\paragraph{Session lessons.} While working on a database, the agent reflects
after each question and records a short conditional rule when something it tried
held, together with the observation that confirmed it. A consolidation step
keeps a compact set per database rather than every candidate.

\paragraph{Cross-session lessons.} The subset of session
lessons that recur across databases, rather than holding only within one, is
promoted to a persistent store and carried into later tasks. Promotion is
outcome-gated: a rule is retained only when later questions continue to bear it
out.

\subsection{Representative rules}

Across the BIRD-Dev databases the promoted rules fall into a few recurring
kinds: aggregating without double-counting across one-to-many joins, choosing
the join path that carries a given attribute, recognizing when a stored value is
already a ratio rather than a percentage, projecting bridge tables without
duplicate rows, and reading compound filter phrasing. Each is a short,
human-readable conditional paired with the evidence that confirmed it.
Table~\ref{tab:appB-rules} gives representative rules with that evidence.

\begin{table}[ht]
\centering
\footnotesize
\renewcommand{\arraystretch}{1.25}
\begin{tabular}{@{}p{1.8cm} p{6.7cm} p{6.2cm}@{}}
\toprule
\textbf{Kind} & \textbf{Rule} & \textbf{Evidence} \\
\midrule
Output format & Compute a percentage from counts; do not read a similarly named stored column, which often holds a 0--1 ratio. & A stored column held about 0.70 where the question wanted 70.15 computed. \\
Filter value & A strict ``oldest and lowest $X$'' intersection often returns nothing; probe it, then read it as ``oldest within the lowest-$X$ group.'' & The strict conjunction returned 0 rows; the relaxed reading returned the intended single row. \\
Filter value & Probe the stored casing of status or legality enums; many are Title Case, so a lowercase literal silently matches nothing. & Filters passed only as \texttt{'Banned'}/\texttt{'Restricted'}, not the lowercase values a hint used. \\
Projection & Querying IDs from a many-to-many bridge table can return an entity several times; add \texttt{DISTINCT} only when the question asks for unique entities. & Gold returned one entity ID four times, all at the same minimum value. \\
Aggregation & ``Spent the most'' means \texttt{SUM} over transactions, not \texttt{MAX} of a single one, even when a hint says \texttt{MAX}. & Ordering people by \texttt{MAX(cost)} failed; summed spending was the intended measure. \\
Output format & Return only the \texttt{YYYY-MM-DD} part of a datetime column when the question asks for a date. & A full timestamp mismatched the bare date the gold answer expected. \\
Filter value & ``After year $X$'' means $>X$, even when a hint maps it to $=X$. & \texttt{=\,'1990'} returned that year's records rather than later ones. \\
\bottomrule
\end{tabular}
\caption{Representative cross-session rules with the execution observation that
confirmed each. Rules are condensed; each is recorded only after the observation
holds on the database at hand.}
\label{tab:appB-rules}
\end{table}

Because each rule is written by the agent itself and re-checked on the live
database before it can change an answer (Section~\ref{sec:memory}), the store
stays interpretable and auditable: a domain expert can read any entry, see the
evidence behind it, and judge whether it should apply, in the same spirit as the
other artifacts DIA produces.

\subsection{Episodic-to-semantic generalization}

A rule does not begin general. The agent first records a concrete observation on
the database it is working, and that within-database lesson is promoted to a
cross-session rule only when later questions bear out the same pattern.
Table~\ref{tab:appB-lifecycle} shows this episodic-to-semantic step for two
rules: the left column is the originating observation, the right is the
generalized rule the store keeps.

\begin{table}[ht]
\centering
\footnotesize
\renewcommand{\arraystretch}{1.25}
\begin{tabular}{@{}p{7.0cm} p{7.2cm}@{}}
\toprule
\textbf{Within-database observation} & \textbf{Cross-database rule} \\
\midrule
On \texttt{california\_schools}, counting schools through the school-to-\texttt{frpm} join, \texttt{COUNT(*)} returned 9{,}977 for one school because it had 9{,}977 fact rows, while \texttt{COUNT(DISTINCT CDSCode)} returned 1. & When counting entities through a join to a table that stores several rows per entity, count with \texttt{COUNT(DISTINCT entity\_pk)} rather than \texttt{COUNT(*)}. \\
On \texttt{financial}, filtering ``account branch in East Bohemia'' through the client's home district returned the wrong population; the account's own district was meant. & Join through the entity that owns the attribute the question names: the account's geography for ``account branch,'' the client's only for ``client lives in.'' \\
\bottomrule
\end{tabular}
\caption{The episodic-to-semantic step. A rule begins as a concrete observation
on one database and is promoted to a cross-session rule only when it generalizes
beyond that database.}
\label{tab:appB-lifecycle}
\end{table}

\subsection{A learned rule in use}

The tiers are consulted, not only written. The trace below shows a session
lesson, formed on earlier questions of \texttt{thrombosis\_prediction} and
re-checked by a live probe before use, redirecting a later answer from a
per-patient collapse to the per-record projection the question intends. We
include it to illustrate the mechanism, not as a measurement of memory's
aggregate effect.

\begin{tracebox}{A session lesson redirects the output (thrombosis\_prediction)}
QUESTION: for each patient born in 1982, state whether their
albumin (ALB) is within the normal range, 3.5 to 5.5.

SESSION LESSON (formed on earlier questions of this database):
  [JOIN-PATH] The Patient->Laboratory join fans out to many lab
  records per patient, so a per-row label is over records, not
  patients. Probe the join size before choosing output granularity.

UNAIDED ATTEMPT: groups by patient, one label per patient.
  SELECT P.ID, CASE WHEN ... THEN 'normal' ELSE 'abnormal' END
  FROM Patient P LEFT JOIN Laboratory L ON P.ID = L.ID
  WHERE P.Birthday LIKE '1982%'
  GROUP BY P.ID;
  --> collapses the fan-out to one row per patient (wrong shape).

WITH THE LESSON IN CONTEXT: probe the join first.
  COUNT(DISTINCT P.ID) = 1     COUNT(*) = 35
  (one matching patient, thirty-five lab records)
Then label one row per record:
  SELECT IIF(L.ALB BETWEEN 3.5 AND 5.5, 'normal', 'abnormal')
  FROM Patient P JOIN Laboratory L ON P.ID = L.ID
  WHERE STRFTIME('%Y', P.Birthday) = '1982';
  --> 35 rows, one per lab record: matches gold.
\end{tracebox}

\section{Production Deployment}
\label{app:deploy}

DIA is deployed in production for enterprise customers. This appendix illustrates a deployment through one workflow, a nursing-staff analysis, carried out as a single conversation. The domain expert uploads a set of operational data files, and in one continuous thread the three agents work in turn over a shared workspace: each builds on the artifacts the previous one produced, and every artifact, the interpretation, the schema, and the analyses, is retained and remains visible for review. The figures are screenshots from this conversation.

\begin{figure}[H]
  \centering
  \includegraphics[width=\linewidth]{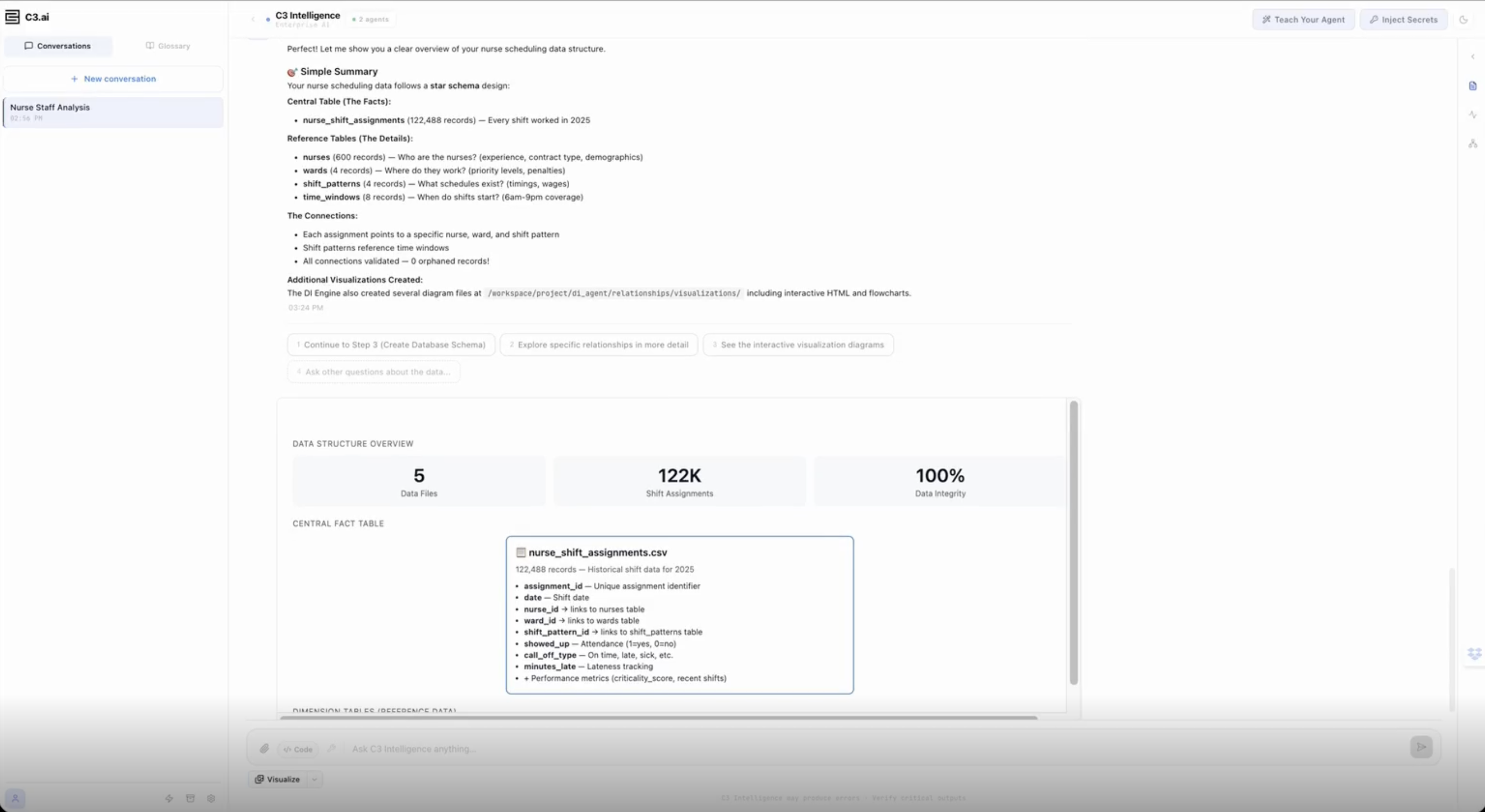}
  \caption{The \emph{Data Interpreter}: the recovered data structure, presented for the domain expert to review.}
  \label{fig:deploy-interp}
\end{figure}

\paragraph{Data Interpreter.}
The \emph{Data Interpreter} inspects the uploaded files and recovers their structure: the entities, the relationships among them, and any data-quality issues. It presents these findings for the domain expert to confirm or correct rather than assuming them (Figure~\ref{fig:deploy-interp}).

\begin{figure}[H]
  \centering
  \includegraphics[width=\linewidth]{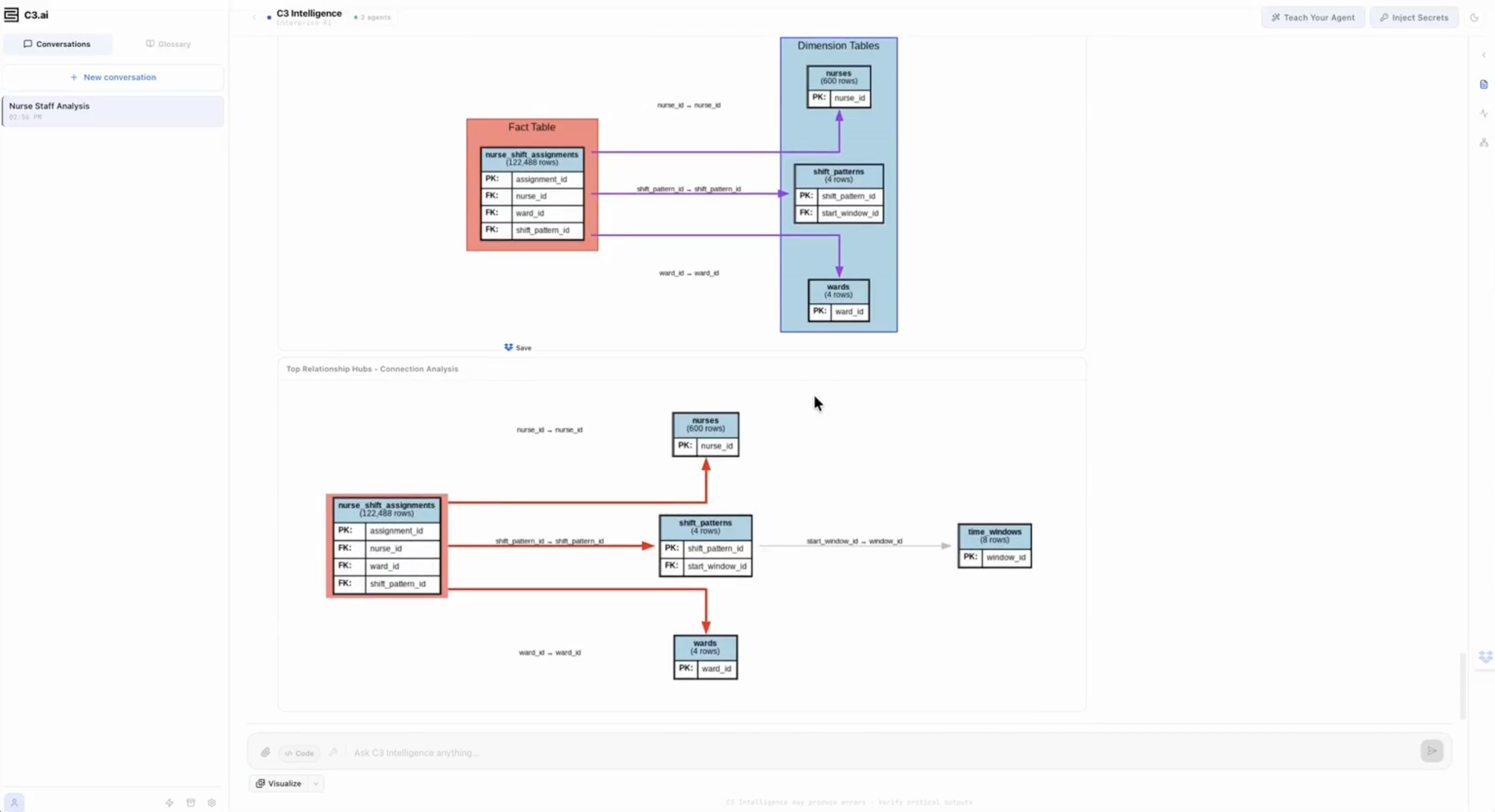}
  \caption{The \emph{Schema Creator}: the resulting database, shown as an entity-relationship diagram.}
  \label{fig:deploy-schema}
\end{figure}

\paragraph{Schema Creator.}
Once the interpretation is confirmed, the \emph{Schema Creator} turns it into a database, declaring the keys and constraints and rendering the result as a schema diagram the expert can inspect (Figure~\ref{fig:deploy-schema}).

\begin{figure}[H]
  \centering
  \includegraphics[width=\linewidth]{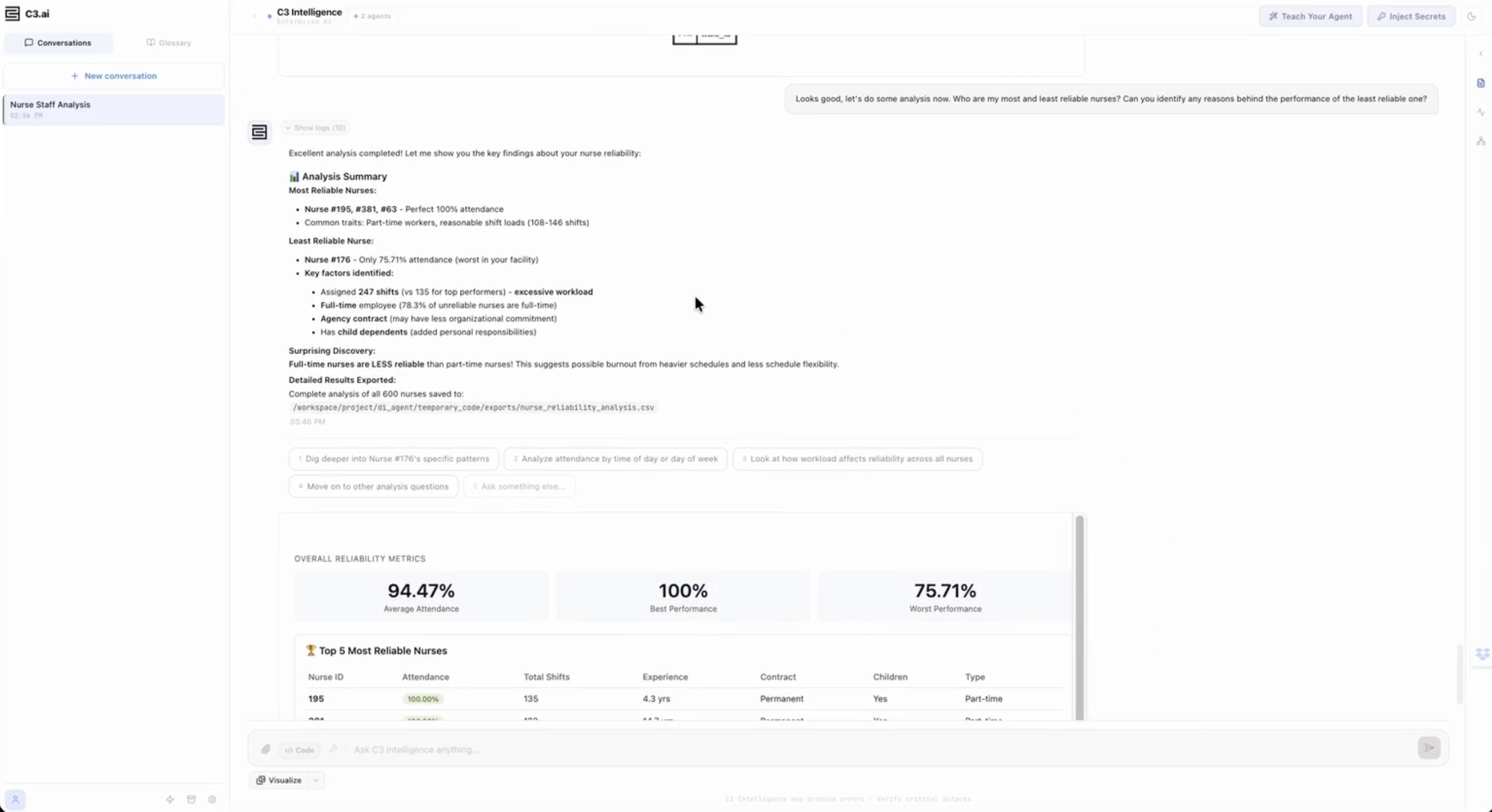}
  \caption{The \emph{Query Generator}: an analytical question answered as a reviewable dashboard, with the steps and exported analysis that produced it.}
  \label{fig:deploy-query}
\end{figure}

\paragraph{Query Generator.}
With the database in place, the domain expert asks analytical questions in natural language. The \emph{Query Generator} answers them by writing and executing the SQL queries each analysis requires, returning the result as a dashboard and an exported file alongside the queries that produced them, so the expert can audit the computation rather than trust it (Figure~\ref{fig:deploy-query}).

Because the work happens in one thread over a shared workspace, the expert never writes SQL or DDL yet sees every artifact each agent produced, can correct any step before the next consumes it, and can ask follow-up questions that build on the work already done. The walkthrough uses uploaded files, but the same workflow runs over enterprise source systems through the underlying data platform, which handles connection, ingestion, access control, and execution at scale. This is the execution-grounded, review-at-each-step design the benchmarks measure in isolation, operating here as one continuous deployment.

\paragraph{Deployment lessons.}
Company policy prevents releasing customer counts, workflow volume, or satisfaction metrics, so we report the recurring qualitative pattern instead. The most consequential corrections happen at the \emph{Data Interpreter} to \emph{Schema Creator} handoff: an inferred relationship, key, or semantic type is what the \emph{Schema Creator} materializes into the database, so a misread there, for instance a foreign key inferred on a plausible but incorrect column, would otherwise propagate into every downstream query that touches it. Because the domain expert reviews the interpretation before the \emph{Schema Creator} consumes it (Figure~\ref{fig:deploy-interp}), these upstream errors are caught and corrected at the point where they are cheapest to fix, rather than surfacing later as a wrong analytical answer. This is the practical reason the pipeline reviews each artifact in turn rather than only the final answer: an error in an early, structural artifact is far more consequential than an error in a late, disposable one, and is far harder to notice once several agents downstream have already built on it.

\paragraph{Governance and safeguards.}
The deployment is governed by the same properties that make it auditable. Every step is a concrete, inspectable artifact, so an analysis can be traced from the natural-language question through the executed SQL to the returned result, and any step can be reproduced or corrected. A domain expert reviews each agent's output before the next agent consumes it, keeping a human in the loop at every stage rather than only at the end. Data access runs through the underlying platform, which enforces connection, authentication, and access control, so agents operate only within a user's existing permissions and read only the sources that user is authorized to see. Because generation is execution-grounded, the system surfaces failed or empty executions rather than silently returning an unverified answer, and the memory it accumulates is re-verified against the live database before it can influence a later result (Appendix~\ref{app:memory}), preventing stale rules from silently propagating. These are operational safeguards observed in deployment rather than formal guarantees; strengthening them, for example with automated policy checks on generated write operations, is ongoing work.

\section{Leaderboard Reference}
\label{app:leaderboards}

\begin{figure*}[ht]
  \centering
  \fbox{\includegraphics[width=0.92\textwidth]{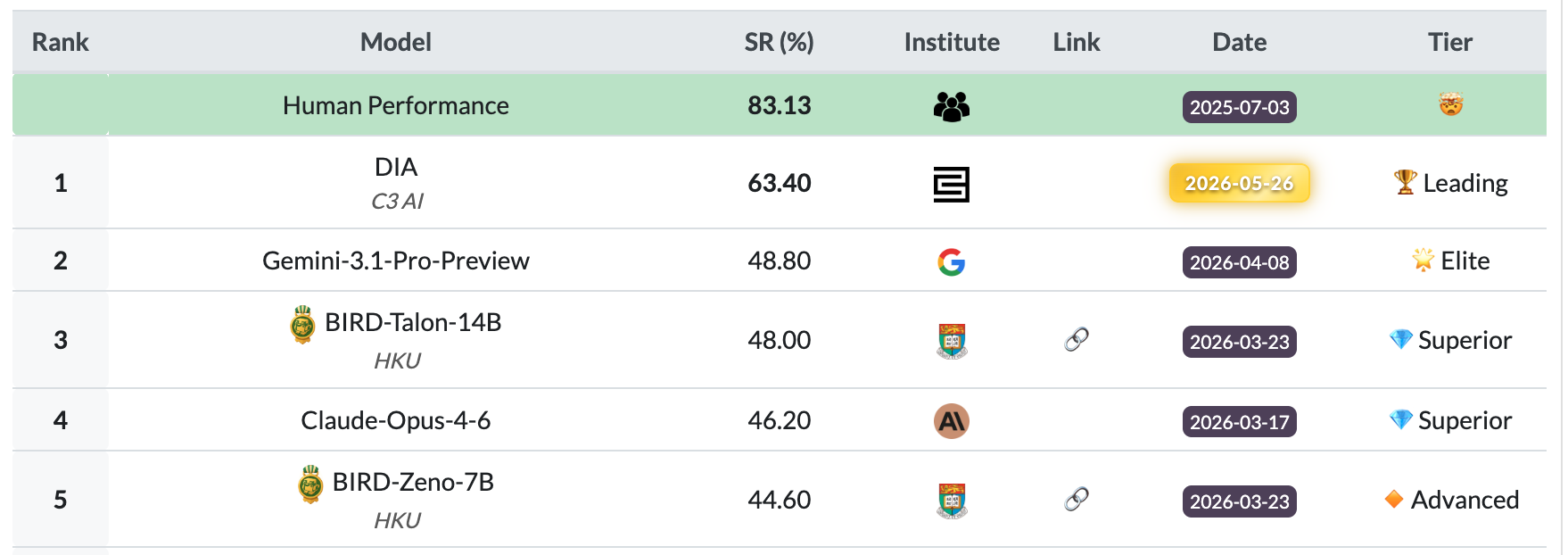}}
  \caption{The BIRD-Critic public leaderboard, BIRD-Critic-SQLite split (\url{https://bird-critic.github.io/}).}
  \label{fig:lb-birdcritic}
\end{figure*}

\begin{figure*}[ht]
  \centering
  \fbox{\includegraphics[width=0.92\textwidth]{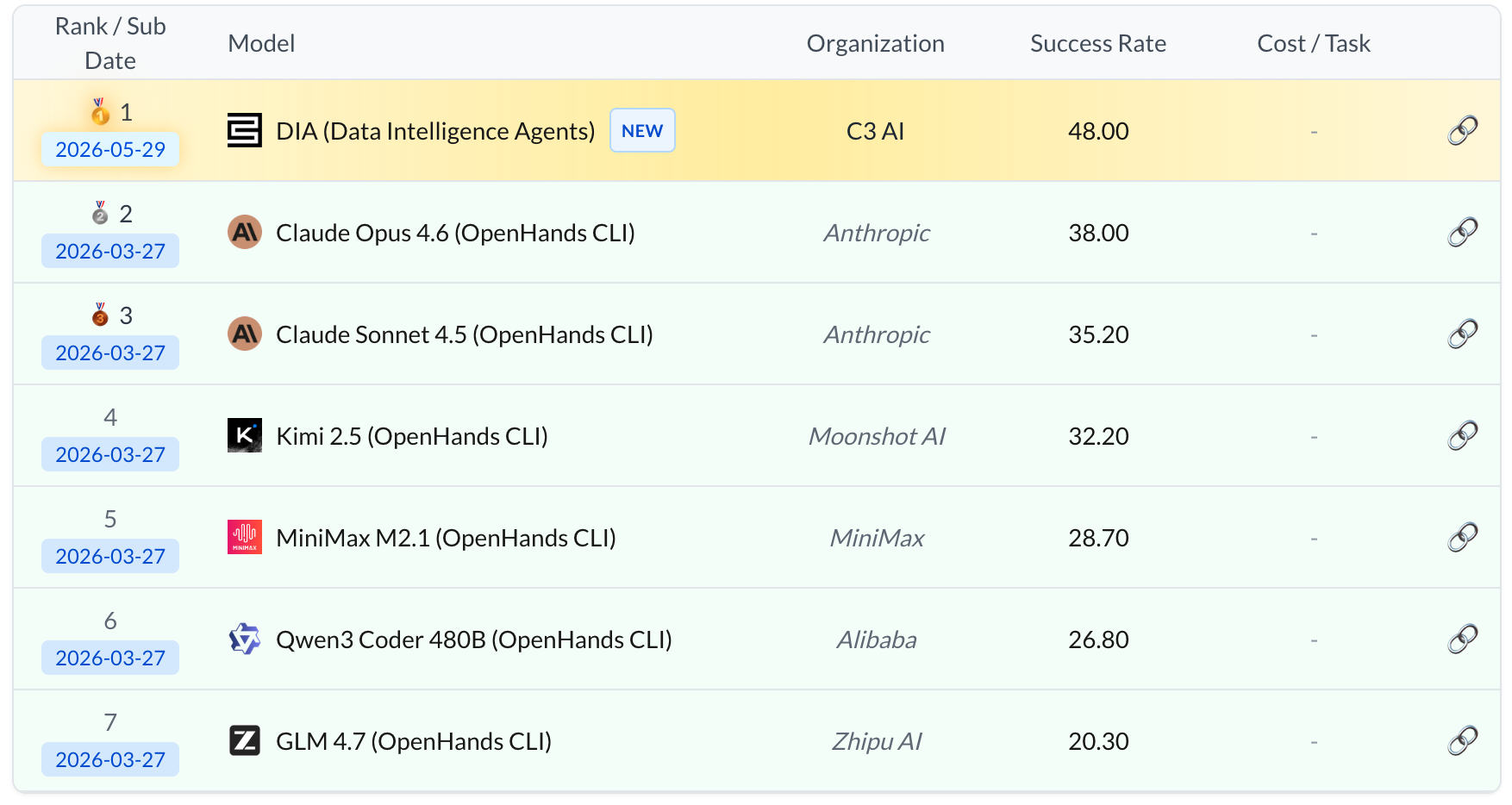}}
  \caption{The LiveSQLBench public leaderboard, LiveSQLBench-Base-Full v1 split (\url{https://livesqlbench.ai/}).}
  \label{fig:lb-livesqlbench}
\end{figure*}

Figures~\ref{fig:lb-birdcritic} and~\ref{fig:lb-livesqlbench} record the public leaderboard standings of BIRD-Critic and LiveSQLBench, from which several of the name-only baselines in Table~\ref{tab:main} are drawn, captured in June 2026.

\section{Configuration}
\label{app:impl}

Table~\ref{tab:config} lists the models and parameters used in our runs. The per-benchmark seed files and prompt templates are described in Appendix~\ref{sec:appA}.

\begin{table}[ht]
\centering
\small
\renewcommand{\arraystretch}{1.15}
\begin{tabular}{@{}l l@{}}
\toprule
\textbf{Parameter} & \textbf{Value} \\
\midrule
Agent framework & OpenHands \\
Agent model & Claude Sonnet 4.5 \\
\midrule
User simulator & o3 \\
Max turns per phase & 15 \\
\midrule
Embedding model & nomic-embed-text-v1.5 \\
Retrieval top-$k$ & 5 \\
Similarity threshold & 0.60 \\
\bottomrule
\end{tabular}
\caption{Experiment configuration. Top to bottom: the shared agent settings; the BIRD-Interact user simulator and per-phase turn cap; and the memory store's embedding model and retrieval parameters (Appendix~\ref{app:memory}).}
\label{tab:config}
\end{table}

\end{document}